\documentclass[%
 aip,
 amsmath,amssymb,
 reprint,%
]{revtex4-2}

\usepackage{graphicx}
\usepackage{dcolumn}
\usepackage{bm}
\usepackage{braket}
\usepackage[utf8]{inputenc}
\usepackage[T1]{fontenc}
\usepackage{mathptmx}
\usepackage{etoolbox}
\usepackage{xcolor}
\makeatletter
\def\@email#1#2{%
 \endgroup
 \patchcmd{\titleblock@produce}
  {\frontmatter@RRAPformat}
  {\frontmatter@RRAPformat{\produce@RRAP{*#1\href{mailto:#2}{#2}}}\frontmatter@RRAPformat}
  {}{}
}%
\makeatother
\begin{document}

\preprint{AIP/123-QED}

\title[Invited perspective]{Nonlinearity and Quantumness in Thermodynamics: From Principles to Technologies}
\author{Gershon Kurizki}
\affiliation{AMOS and Department of Chemical and Biological Physics,
Weizmann Institute of Science, Rehovot 7610001, Israel}
\author{Nilakantha Meher$^*$}
\affiliation{AMOS and Department of Chemical and Biological Physics,
Weizmann Institute of Science, Rehovot 7610001, Israel}
\altaffiliation{AMOS and Department of Chemical and Biological Physics,
Weizmann Institute of Science, Rehovot 7610001, Israel}
\email{nilakantha.meher6@gmail.com}
\affiliation{Department of Physics, SRM University-AP, Amaravati, Andhra Pradesh 522240, India}

\author{Tom\'{a}\v{s} Opatrn\'y}
\affiliation{Department of Optics, Faculty of Science, Palack\'y University, 17. listopadu 50, 77146 Olomouc, Czech Republic}%

\date{\today}

\begin{abstract}
The impact of quantum mechanics on thermodynamics, particularly on the principles and designs of heat machines (HM), has been limited by the incompatibility of quantum coherent evolution with the dissipative, open-system nature of all existing HM and their basic structure, which has not been radically changed since Carnot. We have recently proposed a paradigm change  whereby conventional HM functionality is replaced by that of few-mode coherent, closed systems with nonlinear, e.g. cross-Kerr,  inter-mode couplings. These couplings allow us to coherently filter incident thermal noise, transforming it into a resource of work and information. Current technological advances enable heat engines, noise sensors or microscopes based on such designs to operate with thermal noise sources of few photons. This paradigm shift opens a path towards radically new understanding and exploitation of the relation between coherent, quantum or classical, evolution and thermodynamic behavior. 
\end{abstract}

\maketitle

%

\section{Introduction}
The laws of thermodynamics (TD) emerged in the 19th and early 20th centuries out of a peculiar intertwining of practical and conceptual considerations that shaped the classical framework of the theory. Thus, Carnot’s quest for the ultimate heat engine \cite{CarnotBook} was crystallized by Clausius \cite{clausius1879mechanical} into the fundamental Second Law. Nernst’s strive to understand the limitations of cooling to zero temperature led to the Third law \cite{nernst1907experimental}. It is therefore understandable that quantum mechanics (QM) has long been expected to similarly transform TD, both fundamentally and technologically.  The endeavor to achieve this goal, known as quantum thermodynamics, has been thriving in recent years \cite{kurizki2022Book,BinderBook}.  But has it lived up to the expectations?

\begin{figure}
\includegraphics[scale=0.3]{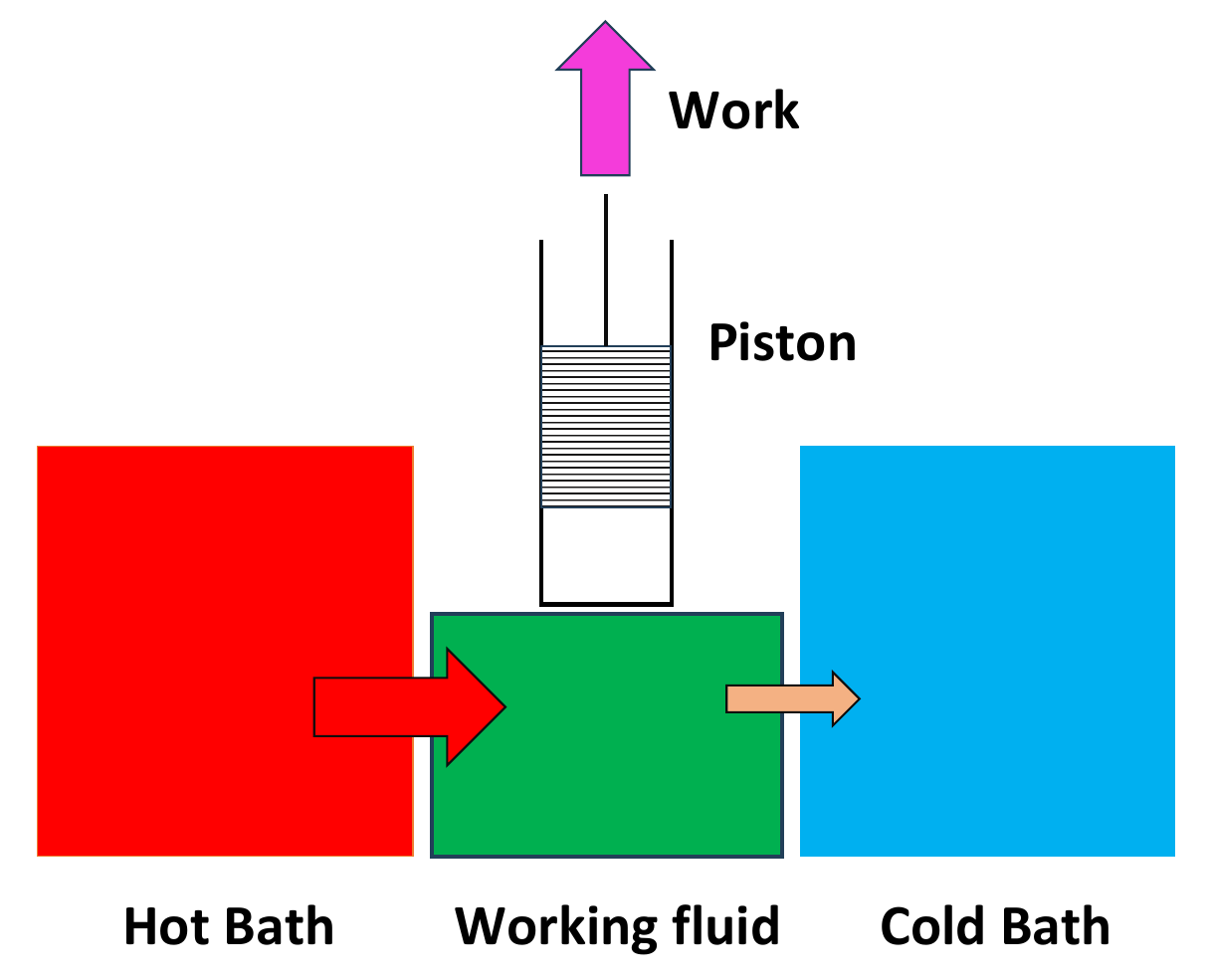}
\caption{A typical heat engine consists of a hot bath, a cold bath and a working fluid. The piston extracts work from the working fluid.}\label{Engine} 
\end{figure}

The rapport of QM and TD  was first brought to the fore by the seminal paper of Scovil and Schulz-DuBois on pumped three-level systems as heat engines (HE) \cite{scovil1959three}. A further boost has been given by the remarkable discovery of Scully et al. that coherently prepared non-thermal baths may yield an effective temperature that allows HE  to surpass the Carnot bound \cite{Scully2003Science,Scully2011PNAS}. Rossnagel et al. \cite{Rossnagel2014PRL} discovered an analogous effect for squeezed baths, notwithstanding the fact that effective temperature is incompatible with the nature of such baths \cite{kurizki2022Book,niedenzu2018quantum}. 

Numerous claims have since been made regarding “quantum advantages” of HE whose working medium and/or energizing baths exhibit quantum   features – quantum  coherence, squeezing,  entanglement \cite{cangemi2024quantum,Ghosh2017,Ghosh2018,ferreri2024quantum,guff2019power,scully2010quantum,scully2011quantum,
hammam2022exploiting,klaers2017squeezed} or multilevel and multipartite collective effects \cite{niedenzu2015performance,niedenzu2018cooperative,tajima2021superconducting,rolandi2023collective,jaramillo2016quantum,
hartmann2020many,zhang2022work,fogarty2020many,souza2022collective,b2020universal}. Yet these claims have to be  critically assessed for two reasons: (i) In most cases,  the claimed quantum advantages have classical counterparts – e.g., squeezing  is not uniquely quantum \cite{loudon1987squeezed,Averbukh1994PRA}, nor is the cooperative response of synchronized dipoles \cite{zhu2015synchronization}.  (ii) \textit{Non-thermal baths}, such as thermal-squeezed or thermal-coherent ones, often endow the working medium with ergotropy (work capacity) along with heat \cite{niedenzu2018quantum}. The effects of nonthermal baths on the engine efficiency bound \cite{niedenzu18quantum} and thermodynamic resources for their preparation \cite{Gardas2015PRE} merit separate considerations, which, however, are not necessarily of quantum nature. \textit{Uniquely quantum advantages} of HE that stem from quantum electrodynamics (QED) are very rare \cite{mukherjee2020anti,xu2022minimal}. 

Has there been a fundamental transformation in the basic concepts of HE owing to the introduction of the above quantum features?  Far from it: all HE since Carnot, be they power plants, aircraft engines or single-atom devices \cite{rossnagel2016single}, have the same basic ingredients: a working medium (WM)/fluid that concentrates the energy from an enormous, macroscopic number of  modes of a  hot bath into a single mode and dumps part of it into a cold bath, converting the difference into work extracted by a piston (Fig. \ref{Engine}). These ingredients are still viewed as essential in any HE. Hence, all HE conceived  thus far, whether  in the classical or  the quantum domain, have been inherently dissipative, open systems \cite{kosloff2014quantum,uzdin2015equivalence,kosloff1984quantum} that do not bode well with quantum unitarity, although coherent effects can play a role in their dynamics.  It should therefore not come as a surprise that QM has not drastically transformed TD, and that its laws, bounds and basic concepts remain essentially unchanged.
  
The pertinent fundamental  question is:  can the long-standing paradigm of TD be transgressed by heat machines (HM), whether HE and refrigerators \cite{cangemi2024quantum}, or thermodynamic batteries \cite{campaioli2024colloquium}, heat diodes and heat transistors \cite{PhysRevE.99.042121,Tahir,segal2005spin,saira2007heat,segal2008single,shen2011single,meher2019atomic,
karimi2017coupled,ronzani2018tunable,joulain2016quantum}?  Namely, can such machines be fully unitary/coherent (non-dissipative) transformers of thermal input channels into non-thermal output channels?

Our response to this question has been given in the approach  we have recently introduced: that of \textit{nonlinear (NL) thermodynamics}. In this approach, multi-channel thermal input noise, which is normally treated as useless, because it contains maximal entropy and cannot produce work (each channel being in a passive state \cite{pusz1978passive,Allahverdyan2004EPL}), is transformed /filtered by a NL element into an output in which some of the channels are in non-passive states. The output channels can serve as thermodynamic resources for work production or sensing.  As a result of such NL filtering, HM can be non-dissipative, fully  coherent devices. These devices must obey the second law of thermodynamics, which forbids entropy decrease of the entire output compared to the entire input, but not entropy redistribution among the channels These NL devices bear analogy to autonomous heat machines \cite{guzman2024key,cangemi2024quantum}, but their operating principles and composition are unconventional.

A caveat is that in order to operate in the quantum domain, one needs \textit{giant NL interactions} that can strongly affect a few-photon field.  Fortuitiously, technology has matured enough to allow for the needed giant NL interactions, particularly in cold gases where photons are converted into Rydberg polaritons \cite{drori2023quantum}. Building on the availability of giant Kerr nonlinearity, we have proposed NL interferometers that can serve as NL coherent HM \cite{Opatrny2023ScAdv}, quantum noise sensors \cite{meher2024thermodynamic} or a supersensitive quantum microscope \cite{meher2024supersensitive}.
Finally, we argue that a “poor man’s alternative” to strong NL interactions, is the effect of quantum measurements  that can similarly steer the dynamics of a  system coupled to thermal modes towards the desired state \cite{OpatrnyPRL21,dasari2022anti}.

\begin{figure}
\includegraphics[scale=0.28]{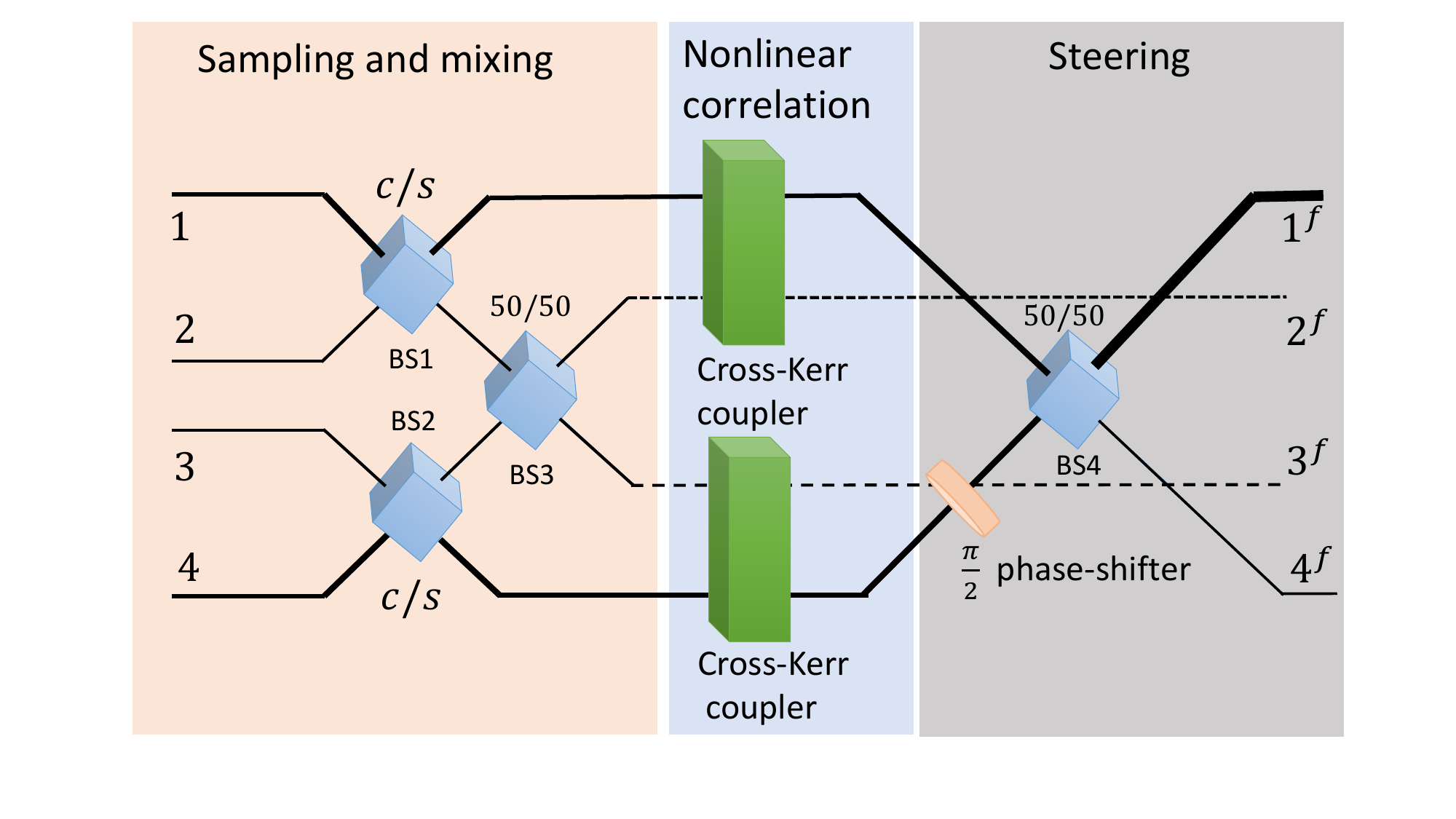}
\caption{Four-mode nonlinear coherent heat engine (based on Ref \cite{Opatrny2023ScAdv}). The engine consists of two hot modes (modes 1 and 4) and two cold modes (modes 2 and 3). For optimal parameter choice, concentration of energy occurs in output mode $1^f$ due to constructive interference between the nonlinearly correlated modes. BS: Beam Splitter}\label{NCHE} 
\end{figure}

\begin{figure}
\includegraphics[scale=0.3]{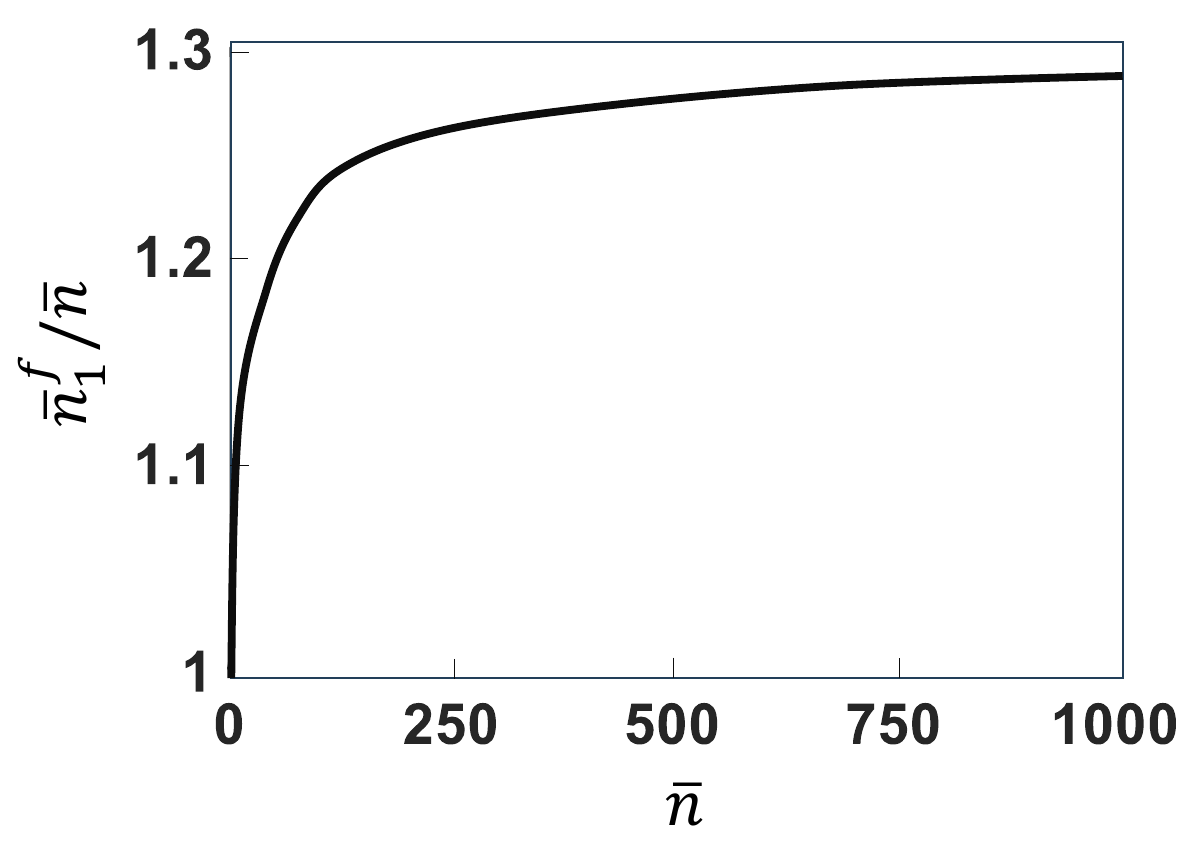}
\caption{Ratio of the mean output photon number in mode $1^f$ to the mean input photon number in mode $1$ in the scheme of Fig. \ref{NCHE}. }\label{NCHEOutput} 
\end{figure}


\section{Coherent nonlinear Heat Engine}
As stated above, we have transgressed the 200-years old TD paradigm by considering unconventional NL, coherent HM comprised of only few hot and cold modes. Such devices do not resort to the usual macroscopic baths, nor to conventional working media. They are instead small, autonomous systems. The energy and entropy/information flow among modes which is required for HM functionality is accomplished in these systems by \textit{nonlinear couplings} between modes. These couplings enable HM operation without intervention  by external control. 
A compact version \cite{Opatrny2023ScAdv} of such a device employs a 4-port interferometer with an input of 2 hot and 2 cold  modes (Fig. \ref{NCHE}), all of them frequency-degenerate.  Linear beam splitters(BS) cannot change their  intensity ratio if the two hot modes have the same temperature, but this ratio becomes controllable when we add two cross-Kerr couplers that correlate or entangle the two pairs of hot and cold modes. The result is  \textit{non-sinusoidal} dependence of the output intensity ratio on  the cross-Kerr coupling strength $\chi$ per photon, which causes the average phase output to depend on $\chi$.

\begin{figure}
\includegraphics[scale=0.5]{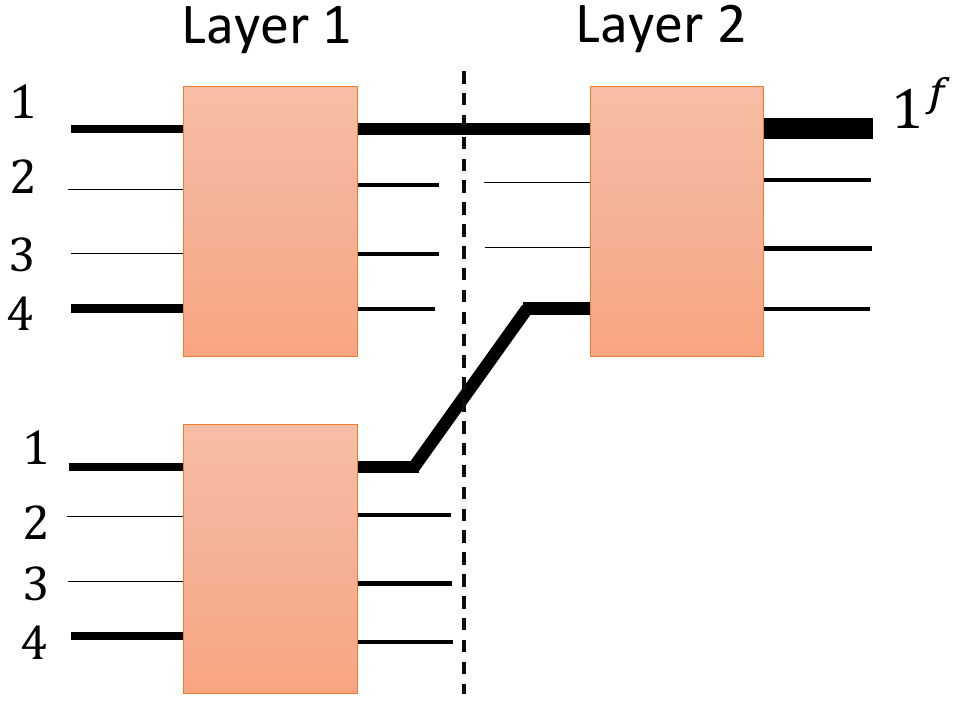}
\caption{A cascade of 4-mode block nonlinear coherent heat machine as in Fig. \ref{NCHE}. Cascade can enhance enhance the concentration of work and energy in output mode $1^f$. }\label{Cascade} 
\end{figure}

In more detail (Fig. \ref{NCHE}), the low-transmission BS1 and BS2 produce weak copies of the hot input modes, then 50:50 BS sample their  random phase difference via the output intensities. NL Kerr cross-couplers correlate those weak copies with the main fractions  of the hot modes such that the phase of each hot mode is shifted depending on the intensity of the other. The phase difference between the output modes depends on the photon-number difference of the weak copies.

Without NL coupling, the random phases of thermal input states wash out the interference between the modes. By contrast, the NL  coupling controls the inter-mode interference, allowing to steer  the energy towards the desired output mode and change the state of the output modes deterministically.  

We can analyze these NL interference effects by treating the thermal input states in the hot modes 1 and 4 as mixtures of coherent states with random mean phases and Gaussian-distributed mean amplitudes. For a given coherent-state, mean intensities  in the output modes $1^f$ and $4^f$, $|\alpha_{1,4}^f|^2$, are related to their input counterparts by the following dependence on the Kerr NL coupling strength $\chi$ (taking the NL coupler length to be 1)
\begin{align}\label{meanoutputcoherent}
|\alpha_{1,4}^f|^2=\frac{c^2}{2}[\alpha_1^2+\alpha_4^2\pm \alpha_1\alpha_4 \sin(2s^2\alpha_1\alpha_4\chi\cos\phi-\phi)],
\end{align} 
where $s^2$ and $c^2$ are the beam-splitter(BS) transmissivity and reflectivity, respectively, in Eq. \eqref{meanoutputcoherent} (see Fig. \ref{NCHE}). This nonsinusoidal dependence of the interference term on the phase difference $\phi$ of the input fields vanishes in the linear limit $\chi=0$, causing interference washout upon averaging over random $\phi$. By contrast, the cross-Kerr nonlinearity allows for output narrow-peaked $\phi-$ phase distribution and yields, for appropriate $\chi$ and $s$,  destructive interference in output mode $4^f$ and constructive interference in output mode $1^f$. It thus steers the mean intensity from mode 4 to mode 1. 

Upon averaging over the random input phase differences $\phi$ and a Gaussian, thermal distribution, the mean intensities of the output hot modes are then given by
\begin{align}
\bar{n}_{1,4}^f=c^2 \bar{n} \left[1\pm \frac{s^2 \chi \bar{n}}{(1+s^4 \chi^2\bar{n}^2)^2} \right],
\end{align}
resulting in energy amplification in output mode $1^f$ (Fig. \ref{NCHEOutput}).

This device adheres to the second law of thermodynamics even though each of the input modes is in a thermal (passive) state, yet some of the output modes are in non-passive states. The consistency with the second law comes about since the input involves hot and cold modes, whose combined state is non-passive. The NL filter concentrates ergotropy in a selected output mode (here $1^f$) at the expense of entropy increase in the other modes.

If we wire many similar 4-mode blocks consecutively, their cascade can realize a multi-mode NL, coherent HM whose work capacity (WC), efficiency  and power scale with the total number of modes (Fig. \ref{Cascade}). The reason is that the output mode states become increasingly non-passive, i.e., their ergotropy/WC grows, when further consecutive blocks, each similar to Fig. \ref{NCHE}, are wired together in a cascade. At any stage $f$ of the cascade, the output WC, $\mathcal{W}$ can be obtained from the sum over quanta-numbers $n$
\begin{align}\label{ergotropydeff}
\mathcal{W}=\hbar\omega\sum_{n}n(p_n^f-p_n^{pas}),
\end{align}
where $\hbar\omega$ is the quantum energy, $p_n^f$ is the $n$-quanta probability and $p_n^{pas}$ its passive-state counterpart that falls off monotonically with $n$, and is obtained by permuting the $n$-quanta components of the state \cite{pusz1978passive,Allahverdyan2004EPL,uzdin2018global,kurizki2022Book}.

The NL inter-mode coupling described above can turn thermal noise into a resource of work and information. Any NL quantum thermodynamic device, including a heat engine, is schematically governed by a product of unitary evolution operators of the form
\begin{align}
\hat U= \hat U_{L1} \hat U_{NL} \hat{U}_{L2}.
\end{align} 
Here $\hat U_{L1}$ and $\hat U_{L2}$ are linear transformation operators that describe the input (sampling and mixing) and output (steering) blocks of the device with the NL correlation operator $\hat U_{NL}$ sandwiched between them (Fig. \ref{NCHE}). While the linear transformation operators can be represented by a sequence of rotations on the Poincar\'e sphere of $(i,j)$ mode-pair Stokes operators that preserve their state Gaussianity, this description does not hold for NL transformations, which can twist, stretch and generally deform the mode-pair observables, and thereby break their input state Gaussianity \cite{Kitagawa1993PRA,Opatrny2015PRA,Opatrny2015PRA2}. The resulting output multi-mode state is then non-Gaussian and quantum-correlated. The phase-space distribution of the output can then be steered to regions of the Poincar\'e sphere where energy is concentrated and work capacity (ergotropy) is non-zero \cite{Opatrny2023ScAdv}.

A simple version of such devices is a nonlinear Mach-Zehnder interferometer (MZI) that correlates two thermal modes by a product of linear transformations (by beam splitters and phase shifters) and NL Kerr cross-coupling \cite{meher2024thermodynamic}. This device \textit{can control the quantum statistics} of each mode: it is a NL noise filter that transforms thermal input  to non-thermal output in a unitary/coherent fashion. Equivalently, such a  device can convert a thermal input state to an output state that is non-passive \cite{meher2024thermodynamic}, i.e. characterized by non-monotonic falloff of the probability as the photon-number grows. Only a nonpassive state can deliver work, which is \textit{an organized form of energy}, by virtue of its ergotropy/work capacity \cite{pusz1978passive,Allahverdyan2004EPL,uzdin2018global,kurizki2022Book}. Thus, nonlinear correlations  can induce such a transformation as an alternative to the conventional working-medium (WM) interactions with heat baths.

\section{Nonlinear Supersensitive Microscopy}
\begin{figure}
\includegraphics[scale=0.4]{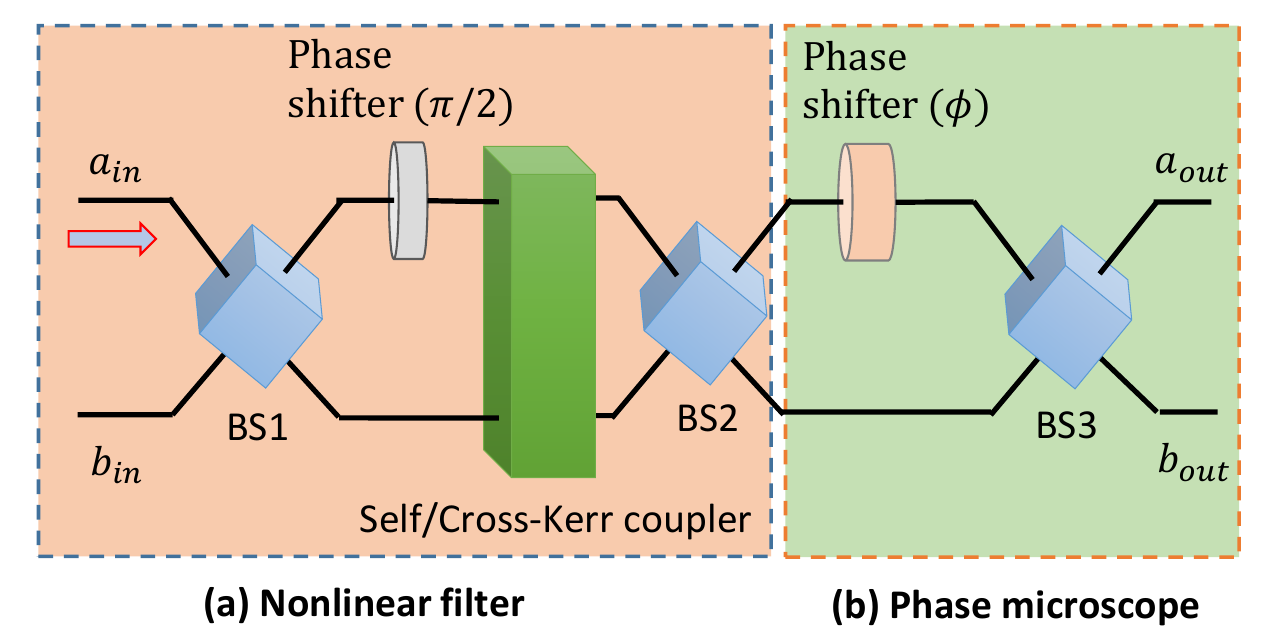}
\caption{Schematic of a (a) nonlinear filter followed by a (b) phase microscope designed to estimate an unknown phase shift (PS) $\phi$. BS: Beam Splitter }\label{PRA1} 
\end{figure}

\begin{figure}
\includegraphics[scale=0.25]{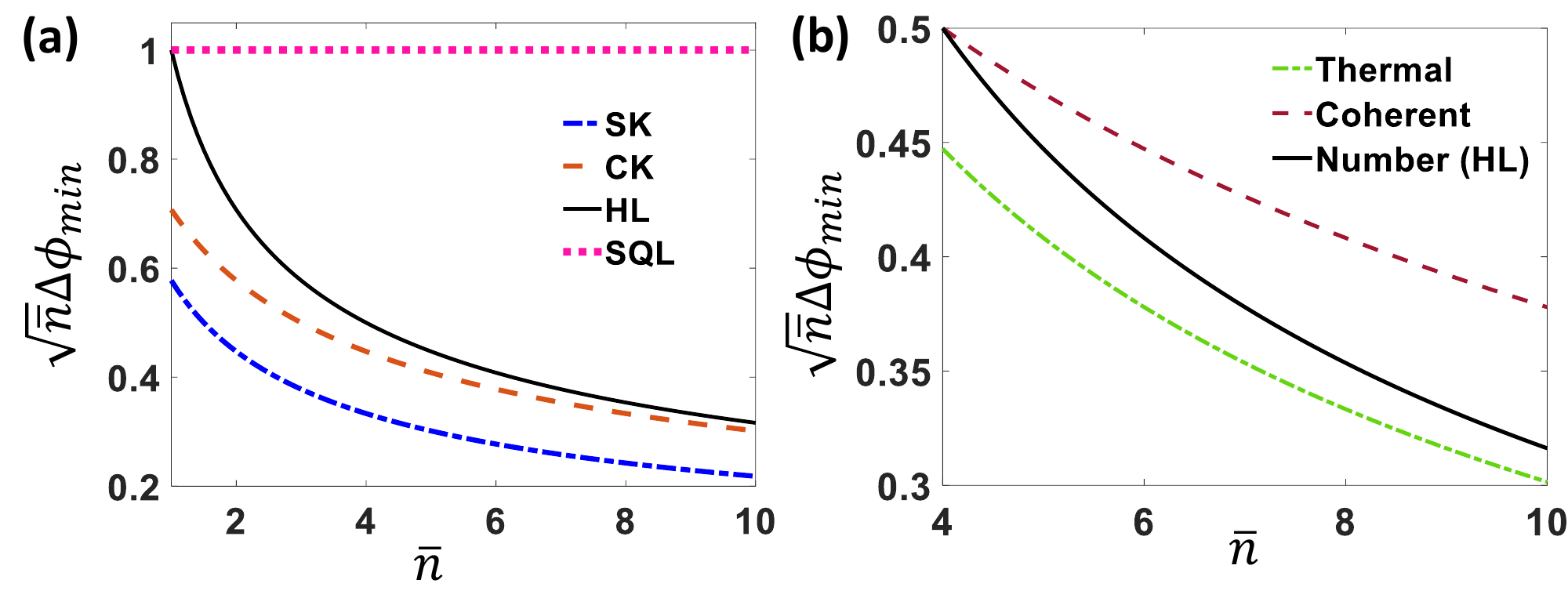}
\caption{ Supersensitive phase estimation (SSPE) in the scheme of Fig. \ref{PRA1}: (a) The normalized minimal phase error $\Delta\phi_{min}$, given by the inverse square root of the quantum Fisher information, as a function of the mean input average photon number $\bar{n}$ for thermal-state input to the interferometer with self-Kerr (SK) or cross-Kerr (CK) nonlinear phase shift per photon $\chi=\pi/2$. The curves indicate phase resolution not only below the standard quantum limit (SQL) but even below Heisenberg limit (HL). (b) The normalized minimal
phase error $\Delta\phi_{min}$ as a function of $\bar{n}$ for thermal, coherent and number-state inputs to the interferometric setup with the same CK nonlinearity as in (a). }\label{PRA2} 
\end{figure}
NL mode couplings  can be used for supersensitive phase estimation (SSPE) microscopy \cite{meher2024supersensitive} in the few-photon domain.  The sensitivity of a transmission microscope realized by an MZI is the minimal size of a sample which we detect via the phase shift of transmitted light $\phi$ through one of the MZI arms. If we shine feeble, classical-like, light the sensitivity or phase resolution is limited by shot noise, corresponding to the standard quantum limit (SQL)\cite{Gerry,Carmichael_BOOK,Gardiner,
ScullyZubairy}: $\Delta\phi \geq 1/\sqrt{\bar{n}}$. Yet the late Y. Silberberg \cite{PhysRevLett.104.123602} at Weizmann showed that the phase resolution  can be brought below the SQL by using N00N two-mode entangled states  $(|n,0\rangle +|0,n\rangle)/\sqrt{2}$.  The SQL limit has also been violated by  other nonclassical (squeezed or entangled coherent) states \cite{lang2013optimal,luis2001equivalence}. These states are however hard to implement and fragile against loss and decoherence, particularly for $n\gg 1$. In contrast to this N00N pure-state input, we have proposed \cite{meher2024supersensitive} to use thermal input, which is usually the worst kind for microscopy, but due to NL filtering (Fig. \ref{PRA1}) the MZI can yield phase resolution not only below the SQL but, surprisingly, even below the Heisenberg limit (HL) \cite{Gerry,Carmichael_BOOK,Gardiner,
ScullyZubairy} $\Delta\phi \leq 1/{\bar{n}}$ (Fig. \ref{PRA2}(a)).

The overall unitary transformation for an MZI with cross-Kerr (CK) nonlinearity that couples modes $a$ and $b$ \cite{chekhova2016nonlinear, kitagawa1986number} is
\begin{subequations}
\begin{align}
\hat U_{CK}&=\hat U_{BS} \hat U_{PS}(\phi) \hat U_{BS} e^{i\chi \hat a^\dagger \hat a \hat b^\dagger \hat b} \hat U_{PS}(\pi/2) \hat U_{BS}.
\end{align}
\end{subequations}
Here, $\hat U_{PS}(\phi)=e^{i\phi \hat a^\dagger \hat a}$ is the operator corresponds to the unknown phase shift $\phi$ in mode $a$, and $\hat U_{BS}$ is the operator of a  50:50 beam splitter (BS) \cite{Gerry}. 

The minimal phase error attainable by a given input state is set by the quantum Cram{\'e}r-Rao bound \cite{cramer1999mathematical} 
\begin{align}\label{CramerRao}
\Delta\phi_{min} \geq \frac{1}{\sqrt{F_Q}}.
\end{align} 
where $F_Q$ is the quantum Fisher information (QFI)\cite{birrittella2021parity}. 

{To evaluate $F_Q$, we need to know the output state after the phase shift operation
\begin{align}
\tilde{\rho}(\phi)=\hat U\rho_{in}\hat U^\dagger,
\end{align}
where 
$\hat U=\hat U_{PS}(\phi) \hat U_{BS} e^{i\chi \hat a^\dagger \hat a \hat b^\dagger \hat b} \hat U_{PS}(\pi/2) \hat U_{BS}$ is the unitary evolution operator of the interferometer up to the phase shifter. {For $\tilde{\rho}(\phi)$, the QFI is given by the formula \cite{birrittella2021parity,Toth2014JPA}
\begin{align}\label{FisherInf}
F_Q(\tilde{\rho})=2\sum_{\substack{k,l \\ \lambda_k + \lambda_l > 0}}\frac{(\lambda_k-\lambda_l)^2}{(\lambda_k+\lambda_l)}|\langle k|a^\dagger a|l\rangle|^2,
\end{align}
where $\{\lambda_k, \ket{k}\}$ are the eigenvalues and the corresponding eigenvectors of $\tilde{\rho}(\phi)$.}  We can obtain from Eq. \eqref{FisherInf} the following $F_Q$ for thermal (t), coherent (c) and number (n) states input filtered by CK, which satisfy a key analytical \textit{inequality} for $\bar{n}> 4$ :
\begin{align}
(F_Q)_{t}= \bar{n}^2+\bar{n}  > (F_Q)_{n}=\bar{n}^2 > (F_Q)_{c}=\tfrac{1}{2}\bar{n}^2+2\bar{n}.
\end{align}
Most remarkably, this result indicates that thermal input yields, under the NL(CK) transformation, the highest phase sensitivity, which is measured by the quantum Fisher information (QFI): it is higher than the phase sensitivity obtained under the same NL-transformation by pure photon-number or coherent-state input with the same mean photon number (Fig. \ref{PRA2}(b)). The reason for this apparently paradoxical result is that the broad statistical spread of thermal light is converted by NL to a broader spread of N00N states than its pure-state counterparts, and since the highest N00N state determines the phase sensitivity bound, this spread yields a higher phase resolution. This result apparently defies the Heisenberg uncertainty limit, which is commonly but inaccurately defined by $\Delta\phi\sim 1/\bar{n}$ ~~\cite{Ou1997PRA}, due to the fact that the states with small $\bar{n}$ with large variance $\Delta n>\bar{n}$ may exhibit $\Delta\phi$ below $1/\bar{n}$~~ \cite{Monras2006PRA,Hofman2009PRA,Pinel2013PRA,Boixo2007PRL,Giovannetti2012PRL}. This finding opens a new path to  supersensitive phase microscopy which has been shown to yield high phase resolution using thermal light sources, even in the presence of high losses \cite{meher2024supersensitive}.

\section{Quantum NL noise sensor}
\begin{figure}
\includegraphics[scale=0.4]{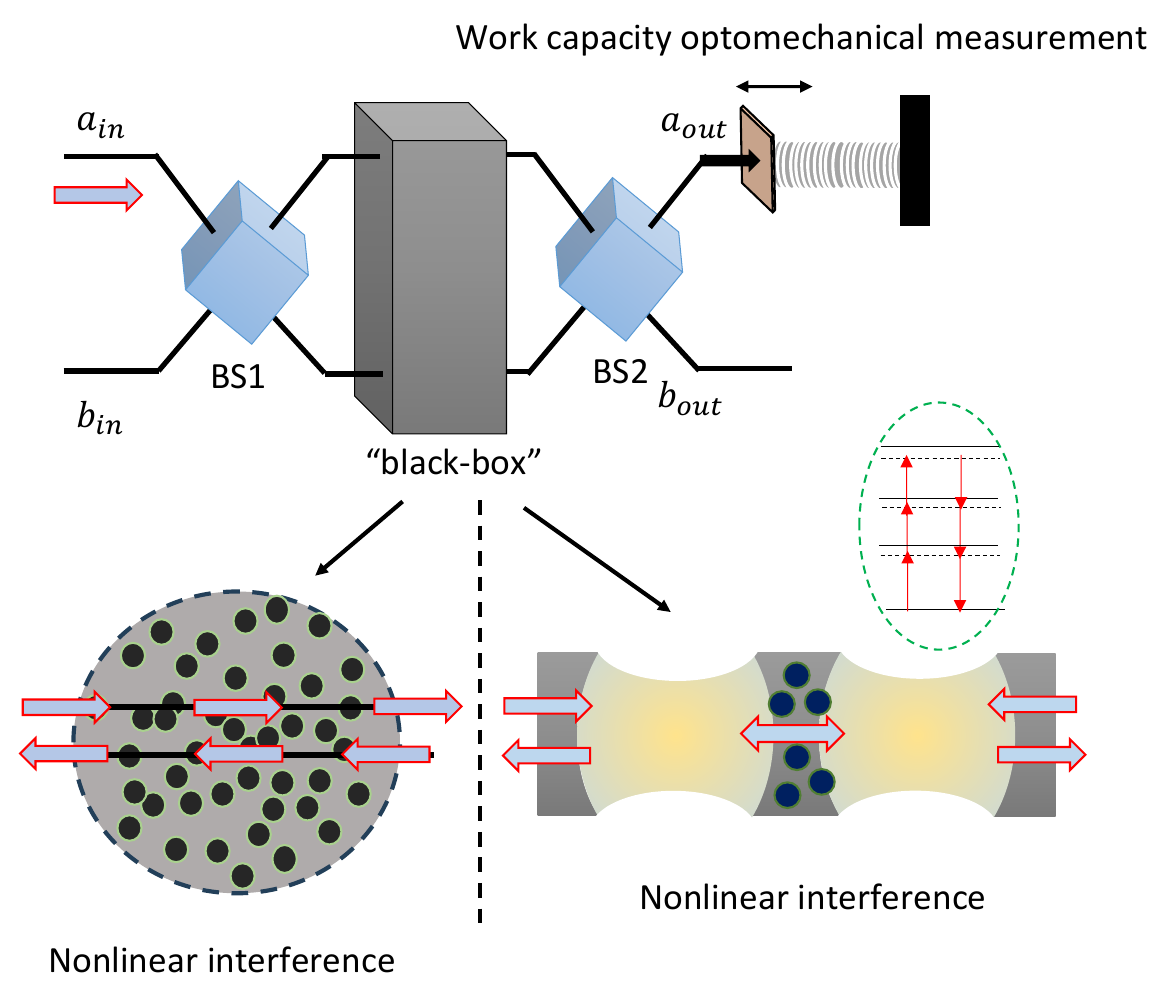}
\caption{Schematic of a Mach-Zehnder interferometer whose internal modes are coupled via an unknown ("black-box") process. This may be either cross-phase dispersive interaction (cold-atom polariton collisions) or multi-photon exchange (cavity modes strongly coupled by atoms, molecules or artificial atoms). The output of the interferometer is coupled to a mechanical oscillator that serves for work capacity measurement. }\label{QST1} 
\end{figure}

\begin{figure}
\includegraphics[scale=0.25]{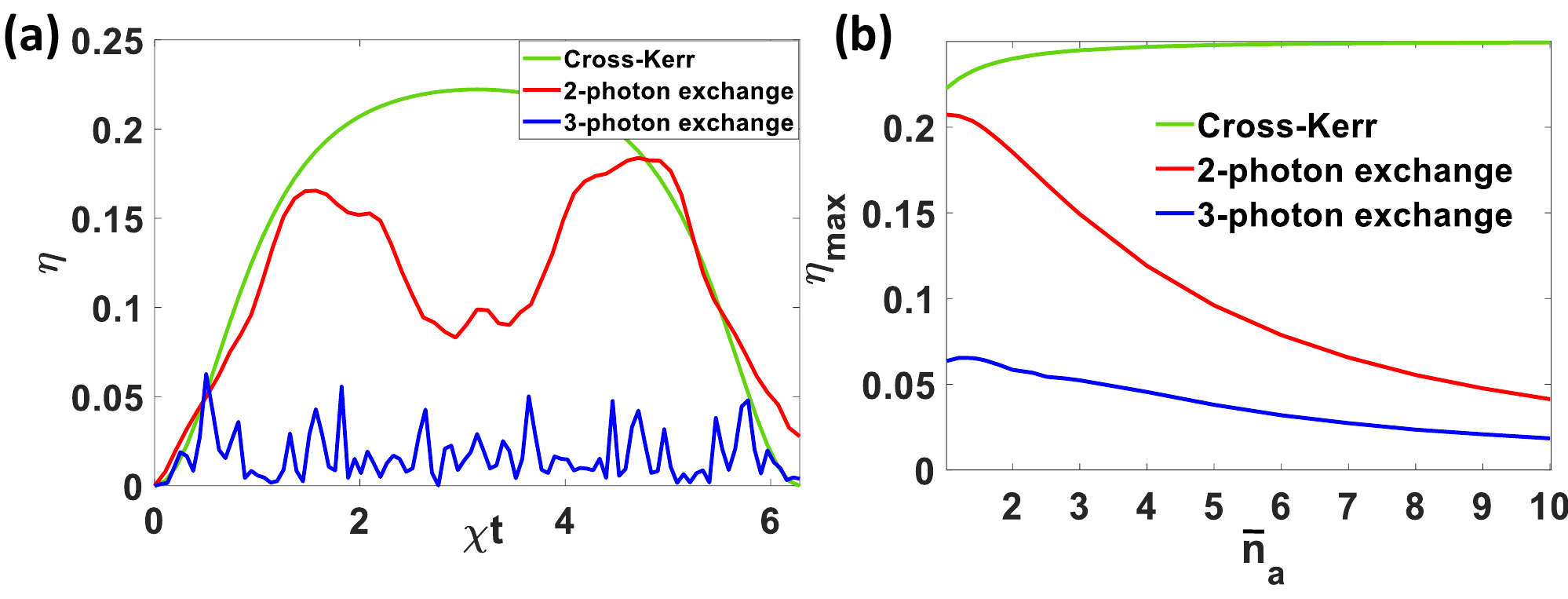}
\caption{(a) Efficiency (the output WC normalized by the input mean photon number) as a function of $\chi t$ for cross-Kerr (solid green), 2-photon exchange (solid red) and 3-photon exchange (solid blue). We consider $\bar{n}_a=1$ and $g=\chi$. (b) Maximal efficiency, denoting the ratio of maximum work capacity generated at the output to the input average photon number in the scheme of Fig. \ref{QST1}, for a "black-box" containing either cross-Kerr, two-photon exchange or three-photon exchange. The saturation behaviour is different for each process, allowing the process to be identified from the output WC.  }\label{QST2} 
\end{figure}

A nonlinear MZI, which endows thermal input with considerable information, can be an altogether different sensor. So far we have only discussed cross-Kerr (CK) MZI, but one may conceive of yet another functionality: suppose that you do not know the Hamiltonian that governs the medium hosted by the MZI and wish to infer it from the output. Such a "black box"  may involve electromagnetic field interactions with multilevel atoms, molecules or impurities in bulk media or in cavities, which may cause NL multi-photon exchange or phase correlations between the two modes of the MZI. Characterization of such processes is of interest for Hamiltonian gauge-field engineering \cite{Chuang1997JModOpt,Goldman2014PRX,del2023dynamical}. Currently, such characterization is  a very hard task, which calls for two-mode quantum tomography \cite{Mohseni2008PRA,Lvovsky2009RMP} that is  much more laborious than classical tomography. By contrast, we  have shown \cite{meher2024thermodynamic} that this characterization can be drastically simplified done by measuring the ergotropy/work capacity (WC) of a single output mode of the MZI: If one output mode is coupled to a mirror suspended on a coherently-driven spring (Fig. \ref{QST1}), then its oscillation energy allows to infer the mean WC of that mode.

The overall goal of this protocol is to identify the particular underlying process in one of the generic categories:\\
\textbf{(a) multi($k$)-photon exchange process}, described by the evolution operator 
\begin{align}\label{kphoton}
\hat U_{NL}^{(k)}=e^{-igt(\hat a^{\dagger k} \hat b^k+\hat a^k \hat b^{\dagger k})},
\end{align}
$g$ being the coupling strength with which the two modes characterized by annihilation operators $\hat a,\hat b$ exchange $k$-quanta,\\
\noindent 
\textbf{(b) nonlinear $s$-order dispersive (cross-phase) coupling}, described by the evolution operator \cite{Carmichael_BOOK,Gardiner,
ScullyZubairy}  that correlates the degenerate two-mode phase shifts
\begin{align}\label{sorderKerr}
\hat U_{NL}^{(s)}(\chi t)=\exp\left[{-i\chi t \hat n_a^s \hat n_b^s}\right].
\end{align} 
Here $\hat n_{a(b)}$ are  the respective number operators of modes $a$ and $b$; {and $\chi$ is the mutual (cross)-Kerr phase-shift when $s\geq 1$ photons in each mode pass  through the NL medium in unit time.} For $s=1$, this evolution operator represents CK coupling.

All of the processes in Eqs. \eqref{kphoton}, \eqref{sorderKerr} transform a Gaussian, passive input state to a non-passive, non-Gaussian output state. The overall two-mode transformation of the input state $\rho_{in}$ to the interferometer output state $\rho_{out}$ is
\begin{subequations} 
\begin{align}
&\rho_{out}=\mathcal{L}_D (\hat U\rho_{in}\hat U^\dagger),\\
&\hat U=\hat U_{L1}\hat U_{NL}\hat U_{L2}
\end{align}
\end{subequations}
where $\mathcal{L}_D$ is the superoperator that accounts for dissipation or decoherence. Under negligible dissipation in the MZI, the resulting mean output WC for {CK or even-$k$ photon exchange processes} can be expressed in terms of the $a$-mode output $n$-photon probabilities $\tilde{P}_n$ as (of Eq. \eqref{ergotropydeff})
\begin{align}\label{ergotropy}
\langle \mathcal{W}_{a}(t) \rangle=\frac{\sum_{n~even} n\tilde{P}_n(t)}{2}=\frac{\langle \hat a^{\dagger}_{out}\hat a_{out}\rangle}{2}.
\end{align}
The WC is thus the energy half of the output average photon number in mode $a$, whereas the other half yields only entropy and hence, amounts to heat production. 

Although expression (\ref{ergotropy}) describes both CK and even-$k$ photon exchange, their WC varies \textit{differently in time} (Fig. \ref{QST2}a), and is thus process-specific \cite{meher2024thermodynamic}. The same is true for {odd-$k$ ($k\geq 3$) photon exchange}, which transforms thermal input into an output with non-zero WC with distinct interaction-time dependence: the WC is found to \textit{oscillate more rapidly} as $k$ increases, due to the increased rate of energy exchange between the modes (Fig. \ref{QST2}a).\\  
\noindent

Cross-phase dispersive coupling, such as CK, and direct $k$-photon exchange differ in the time-dependence and  the \textit{maximum value} of the \textbf{WC efficiency}, i.e., the mean output WC normalized by the mean input energy
\begin{align}
\eta(t)=\frac{\langle \mathcal{W}_{a}(t)\rangle}{\bar{n}_a}.
\end{align}  
For thermal input in mode $a$ and vacuum input in mode $b$, 
the \textit{maximum WC efficiency} of the CK process saturates to $\eta^{CK}_{max}=1/4$ in the classical limit $\bar{n}_a\gg 1$, when the maximal efficiency for direct photon-exchange processes is inversely proportional to the average input photon number in the same limit (Fig. \ref{QST2}b).
Thus, the \textit{output WC efficiency senses the quantum nonlinear process} that the two modes undergo. A linear transformed thermal input does not contribute to WC, i.e., is filtered out by the device.

The above discussion shows that WC characteristics provide a \textit{specific signature} of each of nonlinear processes surveyed above for a given thermal input.

\section{Quantum measurements as a substitute for NL correlations}
\begin{figure}
\includegraphics[scale=0.25]{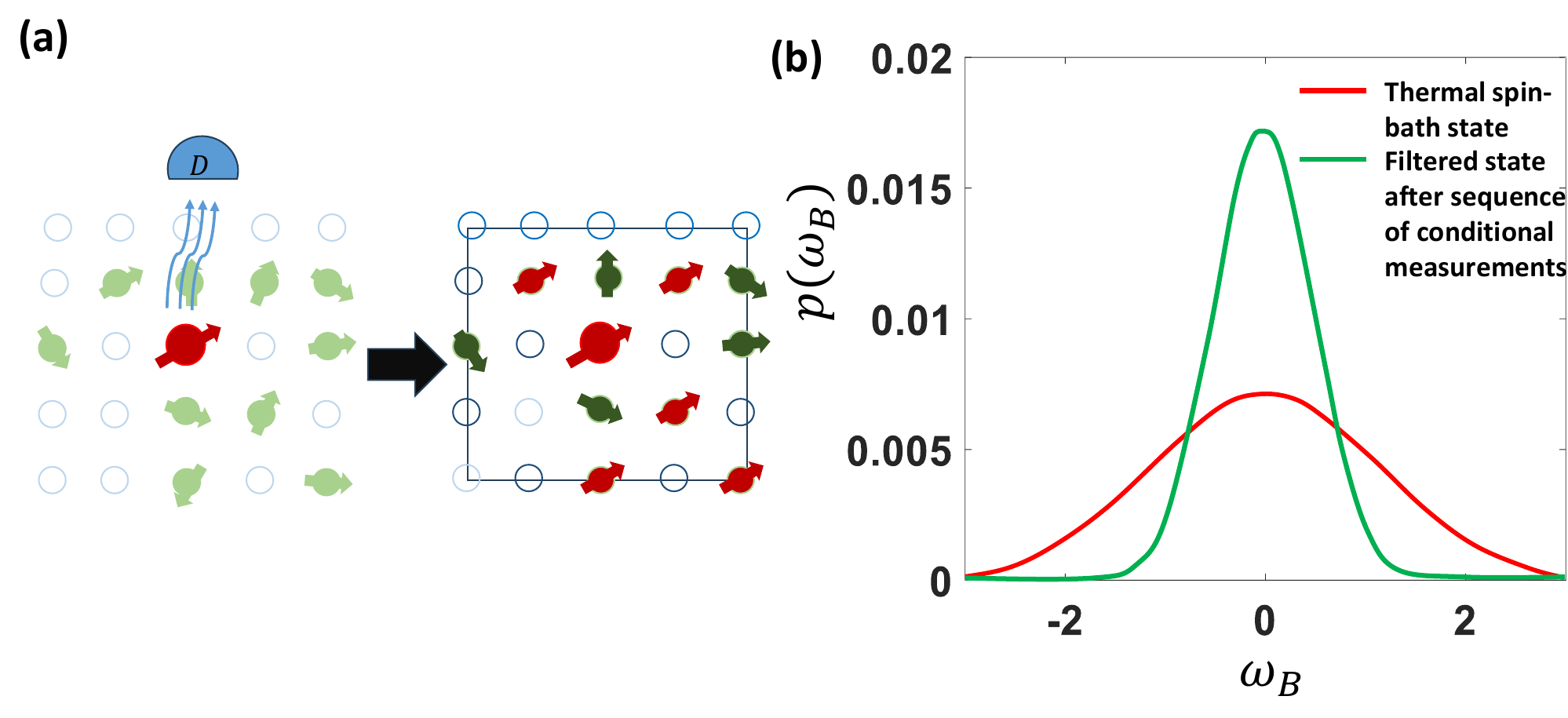}
\caption{(a) Conditional selective measurements of the state of the central probe-spin surrounded by a spin bath in the star configuration. The conditional measurements consist of photon-emission detection from the decayed state or non-detection from the initial state. A sequence of measurements events collapses the spin-bath toward a low entropy state (narrower distribution) with resolvable, partly-(green) or fully-polarized (red) spins. (b) The conditional measurements filter the spin-bath thermal state (red) and render it much narrower (green). } \label{Durga1} 
\end{figure}

Thus far, we have described deterministic, autonomous NL processes as a means of transforming or filtering thermal noise to non-Gaussian states endowed with desired thermodynamic properties: work capacity for heat engines and/or information for metrology or microscopy. However, the engineering of the required NL interactions in the quantum domain is at present feasible only in few experimental setups, particularly in setups involving Rydberg polaritons in cold gases \cite{drori2023quantum} as discussed in Sec. \ref{Conclusion}. Is there a more broadly accessible alternative to such NL processes? 

An alternative is the engineering of desired states of a given system via its measurements by a quantum probe, such as a spin-1/2 or two-level particle, which is coupled to the system by a simple interaction Hamiltonian. A sequence of post-selected events, alias conditional measurements (CM), prepare the desired (target) states at the price of their finite success probability, which can however be optimized.

Early on, CMs were shown to allow the preparation on demand of nonclassical states of a cavity-field mode that is initially prepared in a thermal \cite{harel1996fock} or coherent state \cite{brune1996observing}. This preparation was shown to be achievable by consecutively passing atoms through a high-$Q$ cavity, where the atoms are coupled to the cavity field by the Jaynes-Cummings interaction Hamiltonian \cite{jaynes1963comparison,harel1996optimized} or its off-resonant dispersive \cite{Garraway1994PRA} counterpart upon optimizing the time-intervals between consecutive atoms in the cavity. 

Recently, we have experimentally shown, jointly with our Stuttgart partner, that the filtering of thermal spin-network states into nearly-pure states is feasible by a sequence of few optimized CMs \cite{dasari2022anti}.  The filtering is realizable by frequent measurements  of a probe spin coupled to the spin network initially in a thermal state upon choosing measurement intervals such that the polarization swap of the probe with the network is maximal (Fig. \ref{Durga1}) \cite{dasari2022anti}, thus conforming the anti-Zeno (AZE) regime \cite{kurizki2022Book,kofman2000acceleration,kofman2001universal}. Then, we post-select only those  events where the swap occurred, to which only spins with certain energies contribute. We thereby collapse /filter  the thermal-spin-network  state to a nearly pure collective state. For a nitrogen-vacancy (NV) spin probe in diamond coupled to a network/bath of several (up to 10) nuclear spins, it has been shown \cite{dasari2022anti} that the probe coherence lives 1000 or 10000 times longer in such a filtered network/bath, over $T_1$ time,  compared to the thermal spin network/bath where this time is $T_2$. This purified network/bath state can then serve as an effective sensor or information register. 


There are photonic counterparts to this  probabilistic noise filtering approach, as experimentally demonstrated by our Turin partners based on our old theoretical proposal \cite{VirziPRL2022,kofman2001zeno}.  This experiment has shown that in the anti-Zeno effect (AZE) regime, a sequence of polarization measurements  of a photon can unravel the characteristics of the external noise that affects the photon polarization. Depending upon the polarization measurement outcomes, the photon ensemble is separated into sub-ensembles that exhibit either correlations or anti-correlations between noise-induced phase lumps.

Another measurement-based filtering scheme we have theoretically introduced, employs optimized homodyning, which can transform a thermal input state, which is passive, into a nearly-coherent, non-passive output state with high work capacity \cite{OpatrnyPRL21}.

\section{Conclusions and Outlook}\label{Conclusion}
We have contested the thermodynamic (TD) paradigm that  heat machines (HM) must be open, dissipative systems. Instead,  we have proposed HM to be autonomous, nonlinear  (NL), purely coherent systems.  The simplest  one is the two-mode Mach-Zehnder interferometer (MZI) with cross-Kerr (CK) coupling and its four-mode counterpart. As opposed to linear optical elements, such as beam splitters or phase shifters, NL CK elements allow control or steering of energy flow between hot modes even if they are input at the same temperature \cite{Opatrny2023ScAdv} : NL transformations are required to attain this goal.  A curious exception is a HM energised by a black hole whose gravitational field becomes a heat source for a free fall object, but only  when this field  is reflected by a mirror that orbits the black hole \cite{misra2024black}. Namely, the gravitational field that emanates from the black hole  is in the vacuum state whereas its mirror reflected component has finite temperature.

NL TD schemes may have quantum properties, e.g., in a setup where  each photon can be converted into a cold Rydberg-atom polaritonic excitation so that two photons become cross correlated via long-range dipole-dipole interactions. This amounts to a giant NL CK effect, which was proposed by us \cite{Friedler}, and  has been lately demonstrated by our partners at the Weizmann Institute \cite{drori2023quantum}: An unprecedented  strong CK coupling has been demonstrated by them in a cold rubidium trap where counter-propagating degenerate photonic modes are converted into Rydberg polaritons that induce CK phase shifts of the order of $\pi$ per photon. This finding proves that CK interactions can be  sufficiently large for inducing NL correlations in feeble few-photon thermofields. Atoms confined in high-Q cavities can also yield a single-photon giant Kerr effect \cite{Imamoglu1997PRL}, and so can free electrons in guiding fibers \cite{karnieli2024strong}. Several alternative paths can yield NL TD effects: (i) linear spin-boson interactions that result in a NL transformation of a multi-spin system coupled to a bosonic bath \cite{DBDRao2011PRL}, (ii) post-selected measurement results that mimic NL effects, albeit with limited probability, in photonic or spin-network setups \cite{dasari2022anti,VirziPRL2022}. These alternative schemes can act as NL noise sensors and filters or heat machines on account of the non-Gaussian and non-passive states they can produce. 

In all of the surveyed schemes, the fundamental goal should be to explore the transition from \textit{coherent dynamics to TD} behaviour by scaling up tractable few-mode NL blocks, so as to examine if and when we reach the TD limit.

At the level of applications, we  are exploring a highly  sensitive  tool for  enhancing high-order noise correlations through NL filtering,  either unitarily or by probe measurements, as outlined here. Our hope is to detect  in this way synchronized multi-channel activity in noisy biological systems, which may perhaps hint at  quantum effects.

To conclude, we believe that the surveyed works lay the ground to NL TD and its manifestation in coherent devices that can operate in either classical or the quantum domain.
\section*{Data Availability Statement}
Data sharing is not applicable to this
article as no new data were created or
analyzed in this study.

\section*{AUTHOR DECLARATIONS}
\noindent
\textbf{Conflict of Interest}\\
The authors have no conflicts to disclose.

\section*{References}

\begin{thebibliography}{101}%
\makeatletter
\providecommand \@ifxundefined [1]{%
 \@ifx{#1\undefined}
}%
\providecommand \@ifnum [1]{%
 \ifnum #1\expandafter \@firstoftwo
 \else \expandafter \@secondoftwo
 \fi
}%
\providecommand \@ifx [1]{%
 \ifx #1\expandafter \@firstoftwo
 \else \expandafter \@secondoftwo
 \fi
}%
\providecommand \natexlab [1]{#1}%
\providecommand \enquote  [1]{``#1''}%
\providecommand \bibnamefont  [1]{#1}%
\providecommand \bibfnamefont [1]{#1}%
\providecommand \citenamefont [1]{#1}%
\providecommand \href@noop [0]{\@secondoftwo}%
\providecommand \href [0]{\begingroup \@sanitize@url \@href}%
\providecommand \@href[1]{\@@startlink{#1}\@@href}%
\providecommand \@@href[1]{\endgroup#1\@@endlink}%
\providecommand \@sanitize@url [0]{\catcode `\\12\catcode `\$12\catcode
  `\&12\catcode `\#12\catcode `\^12\catcode `\_12\catcode `\%12\relax}%
\providecommand \@@startlink[1]{}%
\providecommand \@@endlink[0]{}%
\providecommand \url  [0]{\begingroup\@sanitize@url \@url }%
\providecommand \@url [1]{\endgroup\@href {#1}{\urlprefix }}%
\providecommand \urlprefix  [0]{URL }%
\providecommand \Eprint [0]{\href }%
\providecommand \doibase [0]{https://doi.org/}%
\providecommand \selectlanguage [0]{\@gobble}%
\providecommand \bibinfo  [0]{\@secondoftwo}%
\providecommand \bibfield  [0]{\@secondoftwo}%
\providecommand \translation [1]{[#1]}%
\providecommand \BibitemOpen [0]{}%
\providecommand \bibitemStop [0]{}%
\providecommand \bibitemNoStop [0]{.\EOS\space}%
\providecommand \EOS [0]{\spacefactor3000\relax}%
\providecommand \BibitemShut  [1]{\csname bibitem#1\endcsname}%
\let\auto@bib@innerbib\@empty
\bibitem [{\citenamefont {Carnot}(1824)}]{CarnotBook}%
  \BibitemOpen
  \bibfield  {author} {\bibinfo {author} {\bibfnamefont {S.}~\bibnamefont
  {Carnot}},\ }\href@noop {} {\emph {\bibinfo {title} {R\'{e}flexions sur la
  puissance motrice du feu et sur les machines propres \`{a} d\'{e}velopper
  cette puissance}}}\ (\bibinfo  {publisher} {Bachelier Libraire},\ \bibinfo
  {address} {Paris},\ \bibinfo {year} {1824})\BibitemShut {NoStop}%
\bibitem [{\citenamefont {Clausius}(1879)}]{clausius1879mechanical}%
  \BibitemOpen
  \bibfield  {author} {\bibinfo {author} {\bibfnamefont {R.}~\bibnamefont
  {Clausius}},\ }\href@noop {} {\emph {\bibinfo {title} {The mechanical theory
  of heat}}}\ (\bibinfo  {publisher} {Macmillan},\ \bibinfo {year}
  {1879})\BibitemShut {NoStop}%
\bibitem [{\citenamefont {Nernst}(1907)}]{nernst1907experimental}%
  \BibitemOpen
  \bibfield  {author} {\bibinfo {author} {\bibfnamefont {W.}~\bibnamefont
  {Nernst}},\ }\href@noop {} {\emph {\bibinfo {title} {Experimental and
  theoretical applications of thermodynamics to chemistry}}}\ (\bibinfo
  {publisher} {C. Scribner's sons},\ \bibinfo {year} {1907})\BibitemShut
  {NoStop}%
\bibitem [{\citenamefont {Kurizki}\ and\ \citenamefont
  {Kofman}(2022)}]{kurizki2022Book}%
  \BibitemOpen
  \bibfield  {author} {\bibinfo {author} {\bibfnamefont {G.}~\bibnamefont
  {Kurizki}}\ and\ \bibinfo {author} {\bibfnamefont {A.~G.}\ \bibnamefont
  {Kofman}},\ }\href {https://doi.org/10.1017/9781316798454} {\emph {\bibinfo
  {title} {Thermodynamics and Control of Open Quantum Systems}}}\ (\bibinfo
  {publisher} {Cambridge University Press},\ \bibinfo {year}
  {2022})\BibitemShut {NoStop}%
\bibitem [{\citenamefont {Binder}\ \emph {et~al.}(2018)\citenamefont {Binder},
  \citenamefont {Correa}, \citenamefont {Gogolin}, \citenamefont {Anders},\
  and\ \citenamefont {Adesso}}]{BinderBook}%
  \BibitemOpen
  \bibfield  {author} {\bibinfo {author} {\bibfnamefont {F.}~\bibnamefont
  {Binder}}, \bibinfo {author} {\bibfnamefont {L.}~\bibnamefont {Correa}},
  \bibinfo {author} {\bibfnamefont {C.}~\bibnamefont {Gogolin}}, \bibinfo
  {author} {\bibfnamefont {J.}~\bibnamefont {Anders}},\ and\ \bibinfo {author}
  {\bibfnamefont {G.}~\bibnamefont {Adesso}},\ }\href@noop {} {\emph {\bibinfo
  {title} {Thermodynamics in the Quantum Regime: Fundamental Aspects and New
  Directions}}}\ (\bibinfo  {publisher} {Springer},\ \bibinfo {year}
  {2018})\BibitemShut {NoStop}%
\bibitem [{\citenamefont {Scovil}\ and\ \citenamefont
  {Schulz-DuBois}(1959)}]{scovil1959three}%
  \BibitemOpen
  \bibfield  {author} {\bibinfo {author} {\bibfnamefont {H.~E.}\ \bibnamefont
  {Scovil}}\ and\ \bibinfo {author} {\bibfnamefont {E.~O.}\ \bibnamefont
  {Schulz-DuBois}},\ }\bibfield  {title} {\enquote {\bibinfo {title}
  {Three-level masers as heat engines},}\ }\href@noop {} {\bibfield  {journal}
  {\bibinfo  {journal} {Physical Review Letters}\ }\textbf {\bibinfo {volume}
  {2}},\ \bibinfo {pages} {262} (\bibinfo {year} {1959})}\BibitemShut {NoStop}%
\bibitem [{\citenamefont {Scully}\ \emph {et~al.}(2003)\citenamefont {Scully},
  \citenamefont {Zubairy}, \citenamefont {Agarwal},\ and\ \citenamefont
  {Walther}}]{Scully2003Science}%
  \BibitemOpen
  \bibfield  {author} {\bibinfo {author} {\bibfnamefont {M.~O.}\ \bibnamefont
  {Scully}}, \bibinfo {author} {\bibfnamefont {M.~S.}\ \bibnamefont {Zubairy}},
  \bibinfo {author} {\bibfnamefont {G.~S.}\ \bibnamefont {Agarwal}},\ and\
  \bibinfo {author} {\bibfnamefont {H.}~\bibnamefont {Walther}},\ }\bibfield
  {title} {\enquote {\bibinfo {title} {Extracting work from a single heat bath
  via vanishing quantum coherence},}\ }\href
  {https://doi.org/10.1126/science.1078955} {\bibfield  {journal} {\bibinfo
  {journal} {Science}\ }\textbf {\bibinfo {volume} {299}},\ \bibinfo {pages}
  {862--864} (\bibinfo {year} {2003})}\BibitemShut {NoStop}%
\bibitem [{\citenamefont {Scully}\ \emph
  {et~al.}(2011{\natexlab{a}})\citenamefont {Scully}, \citenamefont {Chapin},
  \citenamefont {Dorfman}, \citenamefont {Kim},\ and\ \citenamefont
  {Svidzinsky}}]{Scully2011PNAS}%
  \BibitemOpen
  \bibfield  {author} {\bibinfo {author} {\bibfnamefont {M.~O.}\ \bibnamefont
  {Scully}}, \bibinfo {author} {\bibfnamefont {K.~R.}\ \bibnamefont {Chapin}},
  \bibinfo {author} {\bibfnamefont {K.~E.}\ \bibnamefont {Dorfman}}, \bibinfo
  {author} {\bibfnamefont {M.~B.}\ \bibnamefont {Kim}},\ and\ \bibinfo {author}
  {\bibfnamefont {A.}~\bibnamefont {Svidzinsky}},\ }\bibfield  {title}
  {\enquote {\bibinfo {title} {Quantum heat engine power can be increased by
  noise-induced coherence},}\ }\href {https://doi.org/10.1073/pnas.1110234108}
  {\bibfield  {journal} {\bibinfo  {journal} {Proc. Natl. Acad. Sci.}\ }\textbf
  {\bibinfo {volume} {108}},\ \bibinfo {pages} {15097--15100} (\bibinfo {year}
  {2011}{\natexlab{a}})}\BibitemShut {NoStop}%
\bibitem [{\citenamefont {Ro\ss{}nagel}\ \emph {et~al.}(2014)\citenamefont
  {Ro\ss{}nagel}, \citenamefont {Abah}, \citenamefont {Schmidt-Kaler},
  \citenamefont {Singer},\ and\ \citenamefont {Lutz}}]{Rossnagel2014PRL}%
  \BibitemOpen
  \bibfield  {author} {\bibinfo {author} {\bibfnamefont {J.}~\bibnamefont
  {Ro\ss{}nagel}}, \bibinfo {author} {\bibfnamefont {O.}~\bibnamefont {Abah}},
  \bibinfo {author} {\bibfnamefont {F.}~\bibnamefont {Schmidt-Kaler}}, \bibinfo
  {author} {\bibfnamefont {K.}~\bibnamefont {Singer}},\ and\ \bibinfo {author}
  {\bibfnamefont {E.}~\bibnamefont {Lutz}},\ }\bibfield  {title} {\enquote
  {\bibinfo {title} {Nanoscale heat engine beyond the carnot limit},}\ }\href
  {https://doi.org/10.1103/PhysRevLett.112.030602} {\bibfield  {journal}
  {\bibinfo  {journal} {Phys. Rev. Lett.}\ }\textbf {\bibinfo {volume} {112}},\
  \bibinfo {pages} {030602} (\bibinfo {year} {2014})}\BibitemShut {NoStop}%
\bibitem [{\citenamefont {Niedenzu}\ \emph
  {et~al.}(2018{\natexlab{a}})\citenamefont {Niedenzu}, \citenamefont
  {Mukherjee}, \citenamefont {Ghosh}, \citenamefont {Kofman},\ and\
  \citenamefont {Kurizki}}]{niedenzu2018quantum}%
  \BibitemOpen
  \bibfield  {author} {\bibinfo {author} {\bibfnamefont {W.}~\bibnamefont
  {Niedenzu}}, \bibinfo {author} {\bibfnamefont {V.}~\bibnamefont {Mukherjee}},
  \bibinfo {author} {\bibfnamefont {A.}~\bibnamefont {Ghosh}}, \bibinfo
  {author} {\bibfnamefont {A.~G.}\ \bibnamefont {Kofman}},\ and\ \bibinfo
  {author} {\bibfnamefont {G.}~\bibnamefont {Kurizki}},\ }\bibfield  {title}
  {\enquote {\bibinfo {title} {Quantum engine efficiency bound beyond the
  second law of thermodynamics},}\ }\href@noop {} {\bibfield  {journal}
  {\bibinfo  {journal} {Nature communications}\ }\textbf {\bibinfo {volume}
  {9}},\ \bibinfo {pages} {165} (\bibinfo {year}
  {2018}{\natexlab{a}})}\BibitemShut {NoStop}%
\bibitem [{\citenamefont {Cangemi}, \citenamefont {Bhadra},\ and\ \citenamefont
  {Levy}(2024)}]{cangemi2024quantum}%
  \BibitemOpen
  \bibfield  {author} {\bibinfo {author} {\bibfnamefont {L.~M.}\ \bibnamefont
  {Cangemi}}, \bibinfo {author} {\bibfnamefont {C.}~\bibnamefont {Bhadra}},\
  and\ \bibinfo {author} {\bibfnamefont {A.}~\bibnamefont {Levy}},\ }\bibfield
  {title} {\enquote {\bibinfo {title} {Quantum engines and refrigerators},}\
  }\href@noop {} {\bibfield  {journal} {\bibinfo  {journal} {Physics Reports}\
  }\textbf {\bibinfo {volume} {1087}},\ \bibinfo {pages} {1--71} (\bibinfo
  {year} {2024})}\BibitemShut {NoStop}%
\bibitem [{\citenamefont {Ghosh}\ \emph {et~al.}(2017)\citenamefont {Ghosh},
  \citenamefont {Latune}, \citenamefont {Davidovich},\ and\ \citenamefont
  {Kurizki}}]{Ghosh2017}%
  \BibitemOpen
  \bibfield  {author} {\bibinfo {author} {\bibfnamefont {A.}~\bibnamefont
  {Ghosh}}, \bibinfo {author} {\bibfnamefont {C.~L.}\ \bibnamefont {Latune}},
  \bibinfo {author} {\bibfnamefont {L.}~\bibnamefont {Davidovich}},\ and\
  \bibinfo {author} {\bibfnamefont {G.}~\bibnamefont {Kurizki}},\ }\bibfield
  {title} {\enquote {\bibinfo {title} {Catalysis of heat-to-work conversion in
  quantum machines},}\ }\href {https://doi.org/10.1073/pnas.1711381114}
  {\bibfield  {journal} {\bibinfo  {journal} {Proceedings of the National
  Academy of Sciences}\ }\textbf {\bibinfo {volume} {114}},\ \bibinfo {pages}
  {12156--12161} (\bibinfo {year} {2017})}\BibitemShut {NoStop}%
\bibitem [{\citenamefont {Ghosh}\ \emph {et~al.}(2018)\citenamefont {Ghosh},
  \citenamefont {Gelbwaser-Klimovsky}, \citenamefont {Niedenzu}, \citenamefont
  {Lvovsky}, \citenamefont {Mazets}, \citenamefont {Scully},\ and\
  \citenamefont {Kurizki}}]{Ghosh2018}%
  \BibitemOpen
  \bibfield  {author} {\bibinfo {author} {\bibfnamefont {A.}~\bibnamefont
  {Ghosh}}, \bibinfo {author} {\bibfnamefont {D.}~\bibnamefont
  {Gelbwaser-Klimovsky}}, \bibinfo {author} {\bibfnamefont {W.}~\bibnamefont
  {Niedenzu}}, \bibinfo {author} {\bibfnamefont {A.~I.}\ \bibnamefont
  {Lvovsky}}, \bibinfo {author} {\bibfnamefont {I.}~\bibnamefont {Mazets}},
  \bibinfo {author} {\bibfnamefont {M.~O.}\ \bibnamefont {Scully}},\ and\
  \bibinfo {author} {\bibfnamefont {G.}~\bibnamefont {Kurizki}},\ }\bibfield
  {title} {\enquote {\bibinfo {title} {Two-level masers as heat-to-work
  converters},}\ }\href {https://doi.org/10.1073/pnas.1805354115} {\bibfield
  {journal} {\bibinfo  {journal} {Proceedings of the National Academy of
  Sciences}\ }\textbf {\bibinfo {volume} {115}},\ \bibinfo {pages} {9941--9944}
  (\bibinfo {year} {2018})}\BibitemShut {NoStop}%
\bibitem [{\citenamefont {Ferreri}\ \emph {et~al.}(2024)\citenamefont
  {Ferreri}, \citenamefont {Wang}, \citenamefont {Nori}, \citenamefont
  {Wilhelm},\ and\ \citenamefont {Bruschi}}]{ferreri2024quantum}%
  \BibitemOpen
  \bibfield  {author} {\bibinfo {author} {\bibfnamefont {A.}~\bibnamefont
  {Ferreri}}, \bibinfo {author} {\bibfnamefont {H.}~\bibnamefont {Wang}},
  \bibinfo {author} {\bibfnamefont {F.}~\bibnamefont {Nori}}, \bibinfo {author}
  {\bibfnamefont {F.~K.}\ \bibnamefont {Wilhelm}},\ and\ \bibinfo {author}
  {\bibfnamefont {D.~E.}\ \bibnamefont {Bruschi}},\ }\bibfield  {title}
  {\enquote {\bibinfo {title} {Quantum heat engine based on quantum
  interferometry: the su (1, 1) otto cycle},}\ }\href@noop {} {\bibfield
  {journal} {\bibinfo  {journal} {arXiv preprint arXiv:2409.13411}\ } (\bibinfo
  {year} {2024})}\BibitemShut {NoStop}%
\bibitem [{\citenamefont {Guff}\ \emph {et~al.}(2019)\citenamefont {Guff},
  \citenamefont {Daryanoosh}, \citenamefont {Baragiola},\ and\ \citenamefont
  {Gilchrist}}]{guff2019power}%
  \BibitemOpen
  \bibfield  {author} {\bibinfo {author} {\bibfnamefont {T.}~\bibnamefont
  {Guff}}, \bibinfo {author} {\bibfnamefont {S.}~\bibnamefont {Daryanoosh}},
  \bibinfo {author} {\bibfnamefont {B.~Q.}\ \bibnamefont {Baragiola}},\ and\
  \bibinfo {author} {\bibfnamefont {A.}~\bibnamefont {Gilchrist}},\ }\bibfield
  {title} {\enquote {\bibinfo {title} {Power and efficiency of a thermal engine
  with a coherent bath},}\ }\href@noop {} {\bibfield  {journal} {\bibinfo
  {journal} {Physical Review E}\ }\textbf {\bibinfo {volume} {100}},\ \bibinfo
  {pages} {032129} (\bibinfo {year} {2019})}\BibitemShut {NoStop}%
\bibitem [{\citenamefont {Scully}(2010)}]{scully2010quantum}%
  \BibitemOpen
  \bibfield  {author} {\bibinfo {author} {\bibfnamefont {M.~O.}\ \bibnamefont
  {Scully}},\ }\bibfield  {title} {\enquote {\bibinfo {title} {Quantum
  photocell: Using quantum coherence to reduce radiative recombination<?
  format?> and increase efficiency},}\ }\href@noop {} {\bibfield  {journal}
  {\bibinfo  {journal} {Physical review letters}\ }\textbf {\bibinfo {volume}
  {104}},\ \bibinfo {pages} {207701} (\bibinfo {year} {2010})}\BibitemShut
  {NoStop}%
\bibitem [{\citenamefont {Scully}\ \emph
  {et~al.}(2011{\natexlab{b}})\citenamefont {Scully}, \citenamefont {Chapin},
  \citenamefont {Dorfman}, \citenamefont {Kim},\ and\ \citenamefont
  {Svidzinsky}}]{scully2011quantum}%
  \BibitemOpen
  \bibfield  {author} {\bibinfo {author} {\bibfnamefont {M.~O.}\ \bibnamefont
  {Scully}}, \bibinfo {author} {\bibfnamefont {K.~R.}\ \bibnamefont {Chapin}},
  \bibinfo {author} {\bibfnamefont {K.~E.}\ \bibnamefont {Dorfman}}, \bibinfo
  {author} {\bibfnamefont {M.~B.}\ \bibnamefont {Kim}},\ and\ \bibinfo {author}
  {\bibfnamefont {A.}~\bibnamefont {Svidzinsky}},\ }\bibfield  {title}
  {\enquote {\bibinfo {title} {Quantum heat engine power can be increased by
  noise-induced coherence},}\ }\href@noop {} {\bibfield  {journal} {\bibinfo
  {journal} {Proceedings of the National Academy of Sciences}\ }\textbf
  {\bibinfo {volume} {108}},\ \bibinfo {pages} {15097--15100} (\bibinfo {year}
  {2011}{\natexlab{b}})}\BibitemShut {NoStop}%
\bibitem [{\citenamefont {Hammam}\ \emph {et~al.}(2022)\citenamefont {Hammam},
  \citenamefont {Leitch}, \citenamefont {Hassouni},\ and\ \citenamefont
  {De~Chiara}}]{hammam2022exploiting}%
  \BibitemOpen
  \bibfield  {author} {\bibinfo {author} {\bibfnamefont {K.}~\bibnamefont
  {Hammam}}, \bibinfo {author} {\bibfnamefont {H.}~\bibnamefont {Leitch}},
  \bibinfo {author} {\bibfnamefont {Y.}~\bibnamefont {Hassouni}},\ and\
  \bibinfo {author} {\bibfnamefont {G.}~\bibnamefont {De~Chiara}},\ }\bibfield
  {title} {\enquote {\bibinfo {title} {Exploiting coherence for quantum
  thermodynamic advantage},}\ }\href@noop {} {\bibfield  {journal} {\bibinfo
  {journal} {New Journal of Physics}\ }\textbf {\bibinfo {volume} {24}},\
  \bibinfo {pages} {113053} (\bibinfo {year} {2022})}\BibitemShut {NoStop}%
\bibitem [{\citenamefont {Klaers}\ \emph {et~al.}(2017)\citenamefont {Klaers},
  \citenamefont {Faelt}, \citenamefont {Imamoglu},\ and\ \citenamefont
  {Togan}}]{klaers2017squeezed}%
  \BibitemOpen
  \bibfield  {author} {\bibinfo {author} {\bibfnamefont {J.}~\bibnamefont
  {Klaers}}, \bibinfo {author} {\bibfnamefont {S.}~\bibnamefont {Faelt}},
  \bibinfo {author} {\bibfnamefont {A.}~\bibnamefont {Imamoglu}},\ and\
  \bibinfo {author} {\bibfnamefont {E.}~\bibnamefont {Togan}},\ }\bibfield
  {title} {\enquote {\bibinfo {title} {Squeezed thermal reservoirs as a
  resource for a nanomechanical engine beyond the carnot limit},}\ }\href@noop
  {} {\bibfield  {journal} {\bibinfo  {journal} {Physical Review X}\ }\textbf
  {\bibinfo {volume} {7}},\ \bibinfo {pages} {031044} (\bibinfo {year}
  {2017})}\BibitemShut {NoStop}%
\bibitem [{\citenamefont {Niedenzu}, \citenamefont {Gelbwaser-Klimovsky},\ and\
  \citenamefont {Kurizki}(2015)}]{niedenzu2015performance}%
  \BibitemOpen
  \bibfield  {author} {\bibinfo {author} {\bibfnamefont {W.}~\bibnamefont
  {Niedenzu}}, \bibinfo {author} {\bibfnamefont {D.}~\bibnamefont
  {Gelbwaser-Klimovsky}},\ and\ \bibinfo {author} {\bibfnamefont
  {G.}~\bibnamefont {Kurizki}},\ }\bibfield  {title} {\enquote {\bibinfo
  {title} {Performance limits of multilevel and multipartite quantum heat
  machines},}\ }\href@noop {} {\bibfield  {journal} {\bibinfo  {journal}
  {Physical Review E}\ }\textbf {\bibinfo {volume} {92}},\ \bibinfo {pages}
  {042123} (\bibinfo {year} {2015})}\BibitemShut {NoStop}%
\bibitem [{\citenamefont {Niedenzu}\ and\ \citenamefont
  {Kurizki}(2018)}]{niedenzu2018cooperative}%
  \BibitemOpen
  \bibfield  {author} {\bibinfo {author} {\bibfnamefont {W.}~\bibnamefont
  {Niedenzu}}\ and\ \bibinfo {author} {\bibfnamefont {G.}~\bibnamefont
  {Kurizki}},\ }\bibfield  {title} {\enquote {\bibinfo {title} {Cooperative
  many-body enhancement of quantum thermal machine power},}\ }\href@noop {}
  {\bibfield  {journal} {\bibinfo  {journal} {New Journal of Physics}\ }\textbf
  {\bibinfo {volume} {20}},\ \bibinfo {pages} {113038} (\bibinfo {year}
  {2018})}\BibitemShut {NoStop}%
\bibitem [{\citenamefont {Tajima}\ and\ \citenamefont
  {Funo}(2021)}]{tajima2021superconducting}%
  \BibitemOpen
  \bibfield  {author} {\bibinfo {author} {\bibfnamefont {H.}~\bibnamefont
  {Tajima}}\ and\ \bibinfo {author} {\bibfnamefont {K.}~\bibnamefont {Funo}},\
  }\bibfield  {title} {\enquote {\bibinfo {title} {Superconducting-like heat
  current: Effective cancellation of current-dissipation trade-off by quantum
  coherence},}\ }\href@noop {} {\bibfield  {journal} {\bibinfo  {journal}
  {Physical Review Letters}\ }\textbf {\bibinfo {volume} {127}},\ \bibinfo
  {pages} {190604} (\bibinfo {year} {2021})}\BibitemShut {NoStop}%
\bibitem [{\citenamefont {Rolandi}, \citenamefont {Abiuso},\ and\ \citenamefont
  {Perarnau-Llobet}(2023)}]{rolandi2023collective}%
  \BibitemOpen
  \bibfield  {author} {\bibinfo {author} {\bibfnamefont {A.}~\bibnamefont
  {Rolandi}}, \bibinfo {author} {\bibfnamefont {P.}~\bibnamefont {Abiuso}},\
  and\ \bibinfo {author} {\bibfnamefont {M.}~\bibnamefont {Perarnau-Llobet}},\
  }\bibfield  {title} {\enquote {\bibinfo {title} {Collective advantages in
  finite-time thermodynamics},}\ }\href@noop {} {\bibfield  {journal} {\bibinfo
   {journal} {Physical Review Letters}\ }\textbf {\bibinfo {volume} {131}},\
  \bibinfo {pages} {210401} (\bibinfo {year} {2023})}\BibitemShut {NoStop}%
\bibitem [{\citenamefont {Jaramillo}, \citenamefont {Beau},\ and\ \citenamefont
  {del Campo}(2016)}]{jaramillo2016quantum}%
  \BibitemOpen
  \bibfield  {author} {\bibinfo {author} {\bibfnamefont {J.}~\bibnamefont
  {Jaramillo}}, \bibinfo {author} {\bibfnamefont {M.}~\bibnamefont {Beau}},\
  and\ \bibinfo {author} {\bibfnamefont {A.}~\bibnamefont {del Campo}},\
  }\bibfield  {title} {\enquote {\bibinfo {title} {Quantum supremacy of
  many-particle thermal machines},}\ }\href@noop {} {\bibfield  {journal}
  {\bibinfo  {journal} {New Journal of Physics}\ }\textbf {\bibinfo {volume}
  {18}},\ \bibinfo {pages} {075019} (\bibinfo {year} {2016})}\BibitemShut
  {NoStop}%
\bibitem [{\citenamefont {Hartmann}\ \emph {et~al.}(2020)\citenamefont
  {Hartmann}, \citenamefont {Mukherjee}, \citenamefont {Niedenzu},\ and\
  \citenamefont {Lechner}}]{hartmann2020many}%
  \BibitemOpen
  \bibfield  {author} {\bibinfo {author} {\bibfnamefont {A.}~\bibnamefont
  {Hartmann}}, \bibinfo {author} {\bibfnamefont {V.}~\bibnamefont {Mukherjee}},
  \bibinfo {author} {\bibfnamefont {W.}~\bibnamefont {Niedenzu}},\ and\
  \bibinfo {author} {\bibfnamefont {W.}~\bibnamefont {Lechner}},\ }\bibfield
  {title} {\enquote {\bibinfo {title} {Many-body quantum heat engines with
  shortcuts to adiabaticity},}\ }\href@noop {} {\bibfield  {journal} {\bibinfo
  {journal} {Physical Review Research}\ }\textbf {\bibinfo {volume} {2}},\
  \bibinfo {pages} {023145} (\bibinfo {year} {2020})}\BibitemShut {NoStop}%
\bibitem [{\citenamefont {Zhang}\ and\ \citenamefont
  {Quan}(2022)}]{zhang2022work}%
  \BibitemOpen
  \bibfield  {author} {\bibinfo {author} {\bibfnamefont {F.}~\bibnamefont
  {Zhang}}\ and\ \bibinfo {author} {\bibfnamefont {H.}~\bibnamefont {Quan}},\
  }\bibfield  {title} {\enquote {\bibinfo {title} {Work statistics across a
  quantum critical surface},}\ }\href@noop {} {\bibfield  {journal} {\bibinfo
  {journal} {Physical Review E}\ }\textbf {\bibinfo {volume} {105}},\ \bibinfo
  {pages} {024101} (\bibinfo {year} {2022})}\BibitemShut {NoStop}%
\bibitem [{\citenamefont {Fogarty}\ and\ \citenamefont
  {Busch}(2020)}]{fogarty2020many}%
  \BibitemOpen
  \bibfield  {author} {\bibinfo {author} {\bibfnamefont {T.}~\bibnamefont
  {Fogarty}}\ and\ \bibinfo {author} {\bibfnamefont {T.}~\bibnamefont
  {Busch}},\ }\bibfield  {title} {\enquote {\bibinfo {title} {A many-body heat
  engine at criticality},}\ }\href@noop {} {\bibfield  {journal} {\bibinfo
  {journal} {Quantum Science and Technology}\ }\textbf {\bibinfo {volume}
  {6}},\ \bibinfo {pages} {015003} (\bibinfo {year} {2020})}\BibitemShut
  {NoStop}%
\bibitem [{\citenamefont {Souza}\ \emph {et~al.}(2022)\citenamefont {Souza},
  \citenamefont {Manzano}, \citenamefont {Fazio},\ and\ \citenamefont
  {Iemini}}]{souza2022collective}%
  \BibitemOpen
  \bibfield  {author} {\bibinfo {author} {\bibfnamefont {L.~d.~S.}\
  \bibnamefont {Souza}}, \bibinfo {author} {\bibfnamefont {G.}~\bibnamefont
  {Manzano}}, \bibinfo {author} {\bibfnamefont {R.}~\bibnamefont {Fazio}},\
  and\ \bibinfo {author} {\bibfnamefont {F.}~\bibnamefont {Iemini}},\
  }\bibfield  {title} {\enquote {\bibinfo {title} {Collective effects on the
  performance and stability of quantum heat engines},}\ }\href@noop {}
  {\bibfield  {journal} {\bibinfo  {journal} {Physical Review E}\ }\textbf
  {\bibinfo {volume} {106}},\ \bibinfo {pages} {014143} (\bibinfo {year}
  {2022})}\BibitemShut {NoStop}%
\bibitem [{\citenamefont {B.~S}\ \emph {et~al.}(2020)\citenamefont {B.~S},
  \citenamefont {Mukherjee}, \citenamefont {Divakaran},\ and\ \citenamefont
  {del Campo}}]{b2020universal}%
  \BibitemOpen
  \bibfield  {author} {\bibinfo {author} {\bibfnamefont {R.}~\bibnamefont
  {B.~S}}, \bibinfo {author} {\bibfnamefont {V.}~\bibnamefont {Mukherjee}},
  \bibinfo {author} {\bibfnamefont {U.}~\bibnamefont {Divakaran}},\ and\
  \bibinfo {author} {\bibfnamefont {A.}~\bibnamefont {del Campo}},\ }\bibfield
  {title} {\enquote {\bibinfo {title} {Universal finite-time thermodynamics of
  many-body quantum machines from kibble-zurek scaling},}\ }\href@noop {}
  {\bibfield  {journal} {\bibinfo  {journal} {Physical Review Research}\
  }\textbf {\bibinfo {volume} {2}},\ \bibinfo {pages} {043247} (\bibinfo {year}
  {2020})}\BibitemShut {NoStop}%
\bibitem [{\citenamefont {Loudon}\ and\ \citenamefont
  {Knight}(1987)}]{loudon1987squeezed}%
  \BibitemOpen
  \bibfield  {author} {\bibinfo {author} {\bibfnamefont {R.}~\bibnamefont
  {Loudon}}\ and\ \bibinfo {author} {\bibfnamefont {P.~L.}\ \bibnamefont
  {Knight}},\ }\bibfield  {title} {\enquote {\bibinfo {title} {Squeezed
  light},}\ }\href@noop {} {\bibfield  {journal} {\bibinfo  {journal} {Journal
  of modern optics}\ }\textbf {\bibinfo {volume} {34}},\ \bibinfo {pages}
  {709--759} (\bibinfo {year} {1987})}\BibitemShut {NoStop}%
\bibitem [{\citenamefont {Averbukh}, \citenamefont {Sherman},\ and\
  \citenamefont {Kurizki}(1994)}]{Averbukh1994PRA}%
  \BibitemOpen
  \bibfield  {author} {\bibinfo {author} {\bibfnamefont {I.}~\bibnamefont
  {Averbukh}}, \bibinfo {author} {\bibfnamefont {B.}~\bibnamefont {Sherman}},\
  and\ \bibinfo {author} {\bibfnamefont {G.}~\bibnamefont {Kurizki}},\
  }\bibfield  {title} {\enquote {\bibinfo {title} {Enhanced squeezing by
  periodic frequency modulation under parametric instability conditions},}\
  }\href {https://doi.org/10.1103/PhysRevA.50.5301} {\bibfield  {journal}
  {\bibinfo  {journal} {Phys. Rev. A}\ }\textbf {\bibinfo {volume} {50}},\
  \bibinfo {pages} {5301--5308} (\bibinfo {year} {1994})}\BibitemShut {NoStop}%
\bibitem [{\citenamefont {Zhu}\ \emph {et~al.}(2015)\citenamefont {Zhu},
  \citenamefont {Schachenmayer}, \citenamefont {Xu}, \citenamefont {Herrera},
  \citenamefont {Restrepo}, \citenamefont {Holland},\ and\ \citenamefont
  {Rey}}]{zhu2015synchronization}%
  \BibitemOpen
  \bibfield  {author} {\bibinfo {author} {\bibfnamefont {B.}~\bibnamefont
  {Zhu}}, \bibinfo {author} {\bibfnamefont {J.}~\bibnamefont {Schachenmayer}},
  \bibinfo {author} {\bibfnamefont {M.}~\bibnamefont {Xu}}, \bibinfo {author}
  {\bibfnamefont {F.}~\bibnamefont {Herrera}}, \bibinfo {author} {\bibfnamefont
  {J.~G.}\ \bibnamefont {Restrepo}}, \bibinfo {author} {\bibfnamefont {M.~J.}\
  \bibnamefont {Holland}},\ and\ \bibinfo {author} {\bibfnamefont {A.~M.}\
  \bibnamefont {Rey}},\ }\bibfield  {title} {\enquote {\bibinfo {title}
  {Synchronization of interacting quantum dipoles},}\ }\href@noop {} {\bibfield
   {journal} {\bibinfo  {journal} {New Journal of Physics}\ }\textbf {\bibinfo
  {volume} {17}},\ \bibinfo {pages} {083063} (\bibinfo {year}
  {2015})}\BibitemShut {NoStop}%
\bibitem [{\citenamefont {Niedenzu}\ \emph
  {et~al.}(2018{\natexlab{b}})\citenamefont {Niedenzu}, \citenamefont
  {Mukherjee}, \citenamefont {Ghosh}, \citenamefont {Kofman},\ and\
  \citenamefont {Kurizki}}]{niedenzu18quantum}%
  \BibitemOpen
  \bibfield  {author} {\bibinfo {author} {\bibfnamefont {W.}~\bibnamefont
  {Niedenzu}}, \bibinfo {author} {\bibfnamefont {V.}~\bibnamefont {Mukherjee}},
  \bibinfo {author} {\bibfnamefont {A.}~\bibnamefont {Ghosh}}, \bibinfo
  {author} {\bibfnamefont {A.~G.}\ \bibnamefont {Kofman}},\ and\ \bibinfo
  {author} {\bibfnamefont {G.}~\bibnamefont {Kurizki}},\ }\bibfield  {title}
  {\enquote {\bibinfo {title} {Quantum engine efficiency bound beyond the
  second law of thermodynamics},}\ }\href
  {https://doi.org/10.1038/s41467-017-01991-6} {\bibfield  {journal} {\bibinfo
  {journal} {Nature Communications}\ }\textbf {\bibinfo {volume} {9}},\
  \bibinfo {pages} {165} (\bibinfo {year} {2018}{\natexlab{b}})}\BibitemShut
  {NoStop}%
\bibitem [{\citenamefont {Gardas}\ and\ \citenamefont
  {Deffner}(2015)}]{Gardas2015PRE}%
  \BibitemOpen
  \bibfield  {author} {\bibinfo {author} {\bibfnamefont {B.}~\bibnamefont
  {Gardas}}\ and\ \bibinfo {author} {\bibfnamefont {S.}~\bibnamefont
  {Deffner}},\ }\bibfield  {title} {\enquote {\bibinfo {title} {Thermodynamic
  universality of quantum carnot engines},}\ }\href
  {https://doi.org/10.1103/PhysRevE.92.042126} {\bibfield  {journal} {\bibinfo
  {journal} {Phys. Rev. E}\ }\textbf {\bibinfo {volume} {92}},\ \bibinfo
  {pages} {042126} (\bibinfo {year} {2015})}\BibitemShut {NoStop}%
\bibitem [{\citenamefont {Mukherjee}, \citenamefont {Kofman},\ and\
  \citenamefont {Kurizki}(2020)}]{mukherjee2020anti}%
  \BibitemOpen
  \bibfield  {author} {\bibinfo {author} {\bibfnamefont {V.}~\bibnamefont
  {Mukherjee}}, \bibinfo {author} {\bibfnamefont {A.~G.}\ \bibnamefont
  {Kofman}},\ and\ \bibinfo {author} {\bibfnamefont {G.}~\bibnamefont
  {Kurizki}},\ }\bibfield  {title} {\enquote {\bibinfo {title} {Anti-zeno
  quantum advantage in fast-driven heat machines},}\ }\href@noop {} {\bibfield
  {journal} {\bibinfo  {journal} {Communications Physics}\ }\textbf {\bibinfo
  {volume} {3}},\ \bibinfo {pages} {8} (\bibinfo {year} {2020})}\BibitemShut
  {NoStop}%
\bibitem [{\citenamefont {Xu}\ \emph {et~al.}(2022)\citenamefont {Xu},
  \citenamefont {Stockburger}, \citenamefont {Kurizki},\ and\ \citenamefont
  {Ankerhold}}]{xu2022minimal}%
  \BibitemOpen
  \bibfield  {author} {\bibinfo {author} {\bibfnamefont {M.}~\bibnamefont
  {Xu}}, \bibinfo {author} {\bibfnamefont {J.}~\bibnamefont {Stockburger}},
  \bibinfo {author} {\bibfnamefont {G.}~\bibnamefont {Kurizki}},\ and\ \bibinfo
  {author} {\bibfnamefont {J.}~\bibnamefont {Ankerhold}},\ }\bibfield  {title}
  {\enquote {\bibinfo {title} {Minimal quantum thermal machine in a bandgap
  environment: non-markovian features and anti-zeno advantage},}\ }\href@noop
  {} {\bibfield  {journal} {\bibinfo  {journal} {New journal of physics}\
  }\textbf {\bibinfo {volume} {24}},\ \bibinfo {pages} {035003} (\bibinfo
  {year} {2022})}\BibitemShut {NoStop}%
\bibitem [{\citenamefont {Ro{\ss}nagel}\ \emph {et~al.}(2016)\citenamefont
  {Ro{\ss}nagel}, \citenamefont {Dawkins}, \citenamefont {Tolazzi},
  \citenamefont {Abah}, \citenamefont {Lutz}, \citenamefont {Schmidt-Kaler},\
  and\ \citenamefont {Singer}}]{rossnagel2016single}%
  \BibitemOpen
  \bibfield  {author} {\bibinfo {author} {\bibfnamefont {J.}~\bibnamefont
  {Ro{\ss}nagel}}, \bibinfo {author} {\bibfnamefont {S.~T.}\ \bibnamefont
  {Dawkins}}, \bibinfo {author} {\bibfnamefont {K.~N.}\ \bibnamefont
  {Tolazzi}}, \bibinfo {author} {\bibfnamefont {O.}~\bibnamefont {Abah}},
  \bibinfo {author} {\bibfnamefont {E.}~\bibnamefont {Lutz}}, \bibinfo {author}
  {\bibfnamefont {F.}~\bibnamefont {Schmidt-Kaler}},\ and\ \bibinfo {author}
  {\bibfnamefont {K.}~\bibnamefont {Singer}},\ }\bibfield  {title} {\enquote
  {\bibinfo {title} {A single-atom heat engine},}\ }\href@noop {} {\bibfield
  {journal} {\bibinfo  {journal} {Science}\ }\textbf {\bibinfo {volume}
  {352}},\ \bibinfo {pages} {325--329} (\bibinfo {year} {2016})}\BibitemShut
  {NoStop}%
\bibitem [{\citenamefont {Kosloff}\ and\ \citenamefont
  {Levy}(2014)}]{kosloff2014quantum}%
  \BibitemOpen
  \bibfield  {author} {\bibinfo {author} {\bibfnamefont {R.}~\bibnamefont
  {Kosloff}}\ and\ \bibinfo {author} {\bibfnamefont {A.}~\bibnamefont {Levy}},\
  }\bibfield  {title} {\enquote {\bibinfo {title} {Quantum heat engines and
  refrigerators: Continuous devices},}\ }\href
  {https://doi.org/10.1146/annurev-physchem-040513-103724} {\bibfield
  {journal} {\bibinfo  {journal} {Annu. Rev. Phys. Chem.}\ }\textbf {\bibinfo
  {volume} {65}},\ \bibinfo {pages} {365--393} (\bibinfo {year}
  {2014})}\BibitemShut {NoStop}%
\bibitem [{\citenamefont {Uzdin}, \citenamefont {Levy},\ and\ \citenamefont
  {Kosloff}(2015)}]{uzdin2015equivalence}%
  \BibitemOpen
  \bibfield  {author} {\bibinfo {author} {\bibfnamefont {R.}~\bibnamefont
  {Uzdin}}, \bibinfo {author} {\bibfnamefont {A.}~\bibnamefont {Levy}},\ and\
  \bibinfo {author} {\bibfnamefont {R.}~\bibnamefont {Kosloff}},\ }\bibfield
  {title} {\enquote {\bibinfo {title} {Equivalence of quantum heat machines,
  and quantum-thermodynamic signatures},}\ }\href@noop {} {\bibfield  {journal}
  {\bibinfo  {journal} {Physical Review X}\ }\textbf {\bibinfo {volume} {5}},\
  \bibinfo {pages} {031044} (\bibinfo {year} {2015})}\BibitemShut {NoStop}%
\bibitem [{\citenamefont {Kosloff}(1984)}]{kosloff1984quantum}%
  \BibitemOpen
  \bibfield  {author} {\bibinfo {author} {\bibfnamefont {R.}~\bibnamefont
  {Kosloff}},\ }\bibfield  {title} {\enquote {\bibinfo {title} {A quantum
  mechanical open system as a model of a heat engine},}\ }\href@noop {}
  {\bibfield  {journal} {\bibinfo  {journal} {The Journal of chemical physics}\
  }\textbf {\bibinfo {volume} {80}},\ \bibinfo {pages} {1625--1631} (\bibinfo
  {year} {1984})}\BibitemShut {NoStop}%
\bibitem [{\citenamefont {Campaioli}\ \emph {et~al.}(2024)\citenamefont
  {Campaioli}, \citenamefont {Gherardini}, \citenamefont {Quach}, \citenamefont
  {Polini},\ and\ \citenamefont {Andolina}}]{campaioli2024colloquium}%
  \BibitemOpen
  \bibfield  {author} {\bibinfo {author} {\bibfnamefont {F.}~\bibnamefont
  {Campaioli}}, \bibinfo {author} {\bibfnamefont {S.}~\bibnamefont
  {Gherardini}}, \bibinfo {author} {\bibfnamefont {J.~Q.}\ \bibnamefont
  {Quach}}, \bibinfo {author} {\bibfnamefont {M.}~\bibnamefont {Polini}},\ and\
  \bibinfo {author} {\bibfnamefont {G.~M.}\ \bibnamefont {Andolina}},\
  }\bibfield  {title} {\enquote {\bibinfo {title} {Colloquium: quantum
  batteries},}\ }\href@noop {} {\bibfield  {journal} {\bibinfo  {journal}
  {Reviews of Modern Physics}\ }\textbf {\bibinfo {volume} {96}},\ \bibinfo
  {pages} {031001} (\bibinfo {year} {2024})}\BibitemShut {NoStop}%
\bibitem [{\citenamefont {Kargi}\ \emph {et~al.}(2019)\citenamefont {Kargi},
  \citenamefont {Naseem}, \citenamefont {Opatrn\'y}, \citenamefont
  {M\"ustecapl\ifmmode \imath \else \i \fi{}o\ifmmode~\breve{g}\else
  \u{g}\fi{}lu},\ and\ \citenamefont {Kurizki}}]{PhysRevE.99.042121}%
  \BibitemOpen
  \bibfield  {author} {\bibinfo {author} {\bibfnamefont {C.}~\bibnamefont
  {Kargi}}, \bibinfo {author} {\bibfnamefont {M.~T.}\ \bibnamefont {Naseem}},
  \bibinfo {author} {\bibfnamefont {T.}~\bibnamefont {Opatrn\'y}}, \bibinfo
  {author} {\bibfnamefont {O.~E.}\ \bibnamefont {M\"ustecapl\ifmmode \imath
  \else \i \fi{}o\ifmmode~\breve{g}\else \u{g}\fi{}lu}},\ and\ \bibinfo
  {author} {\bibfnamefont {G.}~\bibnamefont {Kurizki}},\ }\bibfield  {title}
  {\enquote {\bibinfo {title} {Quantum optical two-atom thermal diode},}\
  }\href {https://doi.org/10.1103/PhysRevE.99.042121} {\bibfield  {journal}
  {\bibinfo  {journal} {Phys. Rev. E}\ }\textbf {\bibinfo {volume} {99}},\
  \bibinfo {pages} {042121} (\bibinfo {year} {2019})}\BibitemShut {NoStop}%
\bibitem [{\citenamefont {Naseem}\ \emph {et~al.}(2020)\citenamefont {Naseem},
  \citenamefont {Misra}, \citenamefont {M\"ustecaplio\ifmmode~\breve{g}\else
  \u{g}\fi{}lu},\ and\ \citenamefont {Kurizki}}]{Tahir}%
  \BibitemOpen
  \bibfield  {author} {\bibinfo {author} {\bibfnamefont {M.~T.}\ \bibnamefont
  {Naseem}}, \bibinfo {author} {\bibfnamefont {A.}~\bibnamefont {Misra}},
  \bibinfo {author} {\bibfnamefont {O.~E.}\ \bibnamefont
  {M\"ustecaplio\ifmmode~\breve{g}\else \u{g}\fi{}lu}},\ and\ \bibinfo {author}
  {\bibfnamefont {G.}~\bibnamefont {Kurizki}},\ }\bibfield  {title} {\enquote
  {\bibinfo {title} {Minimal quantum heat manager boosted by bath spectral
  filtering},}\ }\href {https://doi.org/10.1103/PhysRevResearch.2.033285}
  {\bibfield  {journal} {\bibinfo  {journal} {Phys. Rev. Research}\ }\textbf
  {\bibinfo {volume} {2}},\ \bibinfo {pages} {033285} (\bibinfo {year}
  {2020})}\BibitemShut {NoStop}%
\bibitem [{\citenamefont {Segal}\ and\ \citenamefont
  {Nitzan}(2005)}]{segal2005spin}%
  \BibitemOpen
  \bibfield  {author} {\bibinfo {author} {\bibfnamefont {D.}~\bibnamefont
  {Segal}}\ and\ \bibinfo {author} {\bibfnamefont {A.}~\bibnamefont {Nitzan}},\
  }\bibfield  {title} {\enquote {\bibinfo {title} {Spin-boson thermal
  rectifier},}\ }\href@noop {} {\bibfield  {journal} {\bibinfo  {journal}
  {Physical review letters}\ }\textbf {\bibinfo {volume} {94}},\ \bibinfo
  {pages} {034301} (\bibinfo {year} {2005})}\BibitemShut {NoStop}%
\bibitem [{\citenamefont {Saira}\ \emph {et~al.}(2007)\citenamefont {Saira},
  \citenamefont {Meschke}, \citenamefont {Giazotto}, \citenamefont {Savin},
  \citenamefont {M{\"o}tt{\"o}nen},\ and\ \citenamefont
  {Pekola}}]{saira2007heat}%
  \BibitemOpen
  \bibfield  {author} {\bibinfo {author} {\bibfnamefont {O.-P.}\ \bibnamefont
  {Saira}}, \bibinfo {author} {\bibfnamefont {M.}~\bibnamefont {Meschke}},
  \bibinfo {author} {\bibfnamefont {F.}~\bibnamefont {Giazotto}}, \bibinfo
  {author} {\bibfnamefont {A.~M.}\ \bibnamefont {Savin}}, \bibinfo {author}
  {\bibfnamefont {.~f.~M.}\ \bibnamefont {M{\"o}tt{\"o}nen}},\ and\ \bibinfo
  {author} {\bibfnamefont {J.~P.}\ \bibnamefont {Pekola}},\ }\bibfield  {title}
  {\enquote {\bibinfo {title} {Heat transistor: Demonstration of
  gate-controlled electronic refrigeration},}\ }\href@noop {} {\bibfield
  {journal} {\bibinfo  {journal} {Physical review letters}\ }\textbf {\bibinfo
  {volume} {99}},\ \bibinfo {pages} {027203} (\bibinfo {year}
  {2007})}\BibitemShut {NoStop}%
\bibitem [{\citenamefont {Segal}(2008)}]{segal2008single}%
  \BibitemOpen
  \bibfield  {author} {\bibinfo {author} {\bibfnamefont {D.}~\bibnamefont
  {Segal}},\ }\bibfield  {title} {\enquote {\bibinfo {title} {Single mode heat
  rectifier: Controlling energy flow between electronic conductors},}\
  }\href@noop {} {\bibfield  {journal} {\bibinfo  {journal} {Physical review
  letters}\ }\textbf {\bibinfo {volume} {100}},\ \bibinfo {pages} {105901}
  (\bibinfo {year} {2008})}\BibitemShut {NoStop}%
\bibitem [{\citenamefont {Shen}, \citenamefont {Bradford},\ and\ \citenamefont
  {Shen}(2011)}]{shen2011single}%
  \BibitemOpen
  \bibfield  {author} {\bibinfo {author} {\bibfnamefont {Y.}~\bibnamefont
  {Shen}}, \bibinfo {author} {\bibfnamefont {M.}~\bibnamefont {Bradford}},\
  and\ \bibinfo {author} {\bibfnamefont {J.-T.}\ \bibnamefont {Shen}},\
  }\bibfield  {title} {\enquote {\bibinfo {title} {Single-photon diode by
  exploiting the photon polarization in a waveguide},}\ }\href@noop {}
  {\bibfield  {journal} {\bibinfo  {journal} {Physical review letters}\
  }\textbf {\bibinfo {volume} {107}},\ \bibinfo {pages} {173902} (\bibinfo
  {year} {2011})}\BibitemShut {NoStop}%
\bibitem [{\citenamefont {Meher}\ and\ \citenamefont
  {Sivakumar}(2019)}]{meher2019atomic}%
  \BibitemOpen
  \bibfield  {author} {\bibinfo {author} {\bibfnamefont {N.}~\bibnamefont
  {Meher}}\ and\ \bibinfo {author} {\bibfnamefont {S.}~\bibnamefont
  {Sivakumar}},\ }\bibfield  {title} {\enquote {\bibinfo {title} {Atomic switch
  for control of heat transfer in coupled cavities},}\ }\href@noop {}
  {\bibfield  {journal} {\bibinfo  {journal} {Journal of the Optical Society of
  America B}\ }\textbf {\bibinfo {volume} {37}},\ \bibinfo {pages} {138--147}
  (\bibinfo {year} {2019})}\BibitemShut {NoStop}%
\bibitem [{\citenamefont {Karimi}\ \emph {et~al.}(2017)\citenamefont {Karimi},
  \citenamefont {Pekola}, \citenamefont {Campisi},\ and\ \citenamefont
  {Fazio}}]{karimi2017coupled}%
  \BibitemOpen
  \bibfield  {author} {\bibinfo {author} {\bibfnamefont {B.}~\bibnamefont
  {Karimi}}, \bibinfo {author} {\bibfnamefont {J.}~\bibnamefont {Pekola}},
  \bibinfo {author} {\bibfnamefont {M.}~\bibnamefont {Campisi}},\ and\ \bibinfo
  {author} {\bibfnamefont {R.}~\bibnamefont {Fazio}},\ }\bibfield  {title}
  {\enquote {\bibinfo {title} {Coupled qubits as a quantum heat switch},}\
  }\href@noop {} {\bibfield  {journal} {\bibinfo  {journal} {Quantum Science
  and Technology}\ }\textbf {\bibinfo {volume} {2}},\ \bibinfo {pages} {044007}
  (\bibinfo {year} {2017})}\BibitemShut {NoStop}%
\bibitem [{\citenamefont {Ronzani}\ \emph {et~al.}(2018)\citenamefont
  {Ronzani}, \citenamefont {Karimi}, \citenamefont {Senior}, \citenamefont
  {Chang}, \citenamefont {Peltonen}, \citenamefont {Chen},\ and\ \citenamefont
  {Pekola}}]{ronzani2018tunable}%
  \BibitemOpen
  \bibfield  {author} {\bibinfo {author} {\bibfnamefont {A.}~\bibnamefont
  {Ronzani}}, \bibinfo {author} {\bibfnamefont {B.}~\bibnamefont {Karimi}},
  \bibinfo {author} {\bibfnamefont {J.}~\bibnamefont {Senior}}, \bibinfo
  {author} {\bibfnamefont {Y.-C.}\ \bibnamefont {Chang}}, \bibinfo {author}
  {\bibfnamefont {J.~T.}\ \bibnamefont {Peltonen}}, \bibinfo {author}
  {\bibfnamefont {C.}~\bibnamefont {Chen}},\ and\ \bibinfo {author}
  {\bibfnamefont {J.~P.}\ \bibnamefont {Pekola}},\ }\bibfield  {title}
  {\enquote {\bibinfo {title} {Tunable photonic heat transport in a quantum
  heat valve},}\ }\href@noop {} {\bibfield  {journal} {\bibinfo  {journal}
  {Nature Physics}\ }\textbf {\bibinfo {volume} {14}},\ \bibinfo {pages}
  {991--995} (\bibinfo {year} {2018})}\BibitemShut {NoStop}%
\bibitem [{\citenamefont {Joulain}\ \emph {et~al.}(2016)\citenamefont
  {Joulain}, \citenamefont {Drevillon}, \citenamefont {Ezzahri},\ and\
  \citenamefont {Ordonez-Miranda}}]{joulain2016quantum}%
  \BibitemOpen
  \bibfield  {author} {\bibinfo {author} {\bibfnamefont {K.}~\bibnamefont
  {Joulain}}, \bibinfo {author} {\bibfnamefont {J.}~\bibnamefont {Drevillon}},
  \bibinfo {author} {\bibfnamefont {Y.}~\bibnamefont {Ezzahri}},\ and\ \bibinfo
  {author} {\bibfnamefont {J.}~\bibnamefont {Ordonez-Miranda}},\ }\bibfield
  {title} {\enquote {\bibinfo {title} {Quantum thermal transistor},}\
  }\href@noop {} {\bibfield  {journal} {\bibinfo  {journal} {Physical review
  letters}\ }\textbf {\bibinfo {volume} {116}},\ \bibinfo {pages} {200601}
  (\bibinfo {year} {2016})}\BibitemShut {NoStop}%
\bibitem [{\citenamefont {Pusz}\ and\ \citenamefont
  {Woronowicz}(1978)}]{pusz1978passive}%
  \BibitemOpen
  \bibfield  {author} {\bibinfo {author} {\bibfnamefont {W.}~\bibnamefont
  {Pusz}}\ and\ \bibinfo {author} {\bibfnamefont {S.~L.}\ \bibnamefont
  {Woronowicz}},\ }\bibfield  {title} {\enquote {\bibinfo {title} {Passive
  states and kms states for general quantum systems},}\ }\href@noop {}
  {\bibfield  {journal} {\bibinfo  {journal} {Communications in Mathematical
  Physics}\ }\textbf {\bibinfo {volume} {58}},\ \bibinfo {pages} {273--290}
  (\bibinfo {year} {1978})}\BibitemShut {NoStop}%
\bibitem [{\citenamefont {Allahverdyan}, \citenamefont {Balian},\ and\
  \citenamefont {Nieuwenhuizen}(2004)}]{Allahverdyan2004EPL}%
  \BibitemOpen
  \bibfield  {author} {\bibinfo {author} {\bibfnamefont {A.~E.}\ \bibnamefont
  {Allahverdyan}}, \bibinfo {author} {\bibfnamefont {R.}~\bibnamefont
  {Balian}},\ and\ \bibinfo {author} {\bibfnamefont {T.~M.}\ \bibnamefont
  {Nieuwenhuizen}},\ }\bibfield  {title} {\enquote {\bibinfo {title} {Maximal
  work extraction from finite quantum systems},}\ }\href
  {https://doi.org/10.1209/epl/i2004-10101-2} {\bibfield  {journal} {\bibinfo
  {journal} {Europhysics Letters ({EPL})}\ }\textbf {\bibinfo {volume} {67}},\
  \bibinfo {pages} {565--571} (\bibinfo {year} {2004})}\BibitemShut {NoStop}%
\bibitem [{\citenamefont {Guzm{\'a}n}\ \emph {et~al.}(2024)\citenamefont
  {Guzm{\'a}n}, \citenamefont {Erker}, \citenamefont {Gasparinetti},
  \citenamefont {Huber},\ and\ \citenamefont {Halpern}}]{guzman2024key}%
  \BibitemOpen
  \bibfield  {author} {\bibinfo {author} {\bibfnamefont {J.~A.~M.}\
  \bibnamefont {Guzm{\'a}n}}, \bibinfo {author} {\bibfnamefont
  {P.}~\bibnamefont {Erker}}, \bibinfo {author} {\bibfnamefont
  {S.}~\bibnamefont {Gasparinetti}}, \bibinfo {author} {\bibfnamefont
  {M.}~\bibnamefont {Huber}},\ and\ \bibinfo {author} {\bibfnamefont {N.~Y.}\
  \bibnamefont {Halpern}},\ }\bibfield  {title} {\enquote {\bibinfo {title}
  {Key issues review: useful autonomous quantum machines},}\ }\href@noop {}
  {\bibfield  {journal} {\bibinfo  {journal} {Reports on Progress in Physics}\
  }\textbf {\bibinfo {volume} {87}},\ \bibinfo {pages} {122001} (\bibinfo
  {year} {2024})}\BibitemShut {NoStop}%
\bibitem [{\citenamefont {Drori}\ \emph {et~al.}(2023)\citenamefont {Drori},
  \citenamefont {Das}, \citenamefont {Zohar}, \citenamefont {Winer},
  \citenamefont {Poem}, \citenamefont {Poddubny},\ and\ \citenamefont
  {Firstenberg}}]{drori2023quantum}%
  \BibitemOpen
  \bibfield  {author} {\bibinfo {author} {\bibfnamefont {L.}~\bibnamefont
  {Drori}}, \bibinfo {author} {\bibfnamefont {B.~C.}\ \bibnamefont {Das}},
  \bibinfo {author} {\bibfnamefont {T.~D.}\ \bibnamefont {Zohar}}, \bibinfo
  {author} {\bibfnamefont {G.}~\bibnamefont {Winer}}, \bibinfo {author}
  {\bibfnamefont {E.}~\bibnamefont {Poem}}, \bibinfo {author} {\bibfnamefont
  {A.}~\bibnamefont {Poddubny}},\ and\ \bibinfo {author} {\bibfnamefont
  {O.}~\bibnamefont {Firstenberg}},\ }\bibfield  {title} {\enquote {\bibinfo
  {title} {Quantum vortices of strongly interacting photons},}\ }\href@noop {}
  {\bibfield  {journal} {\bibinfo  {journal} {Science}\ }\textbf {\bibinfo
  {volume} {381}},\ \bibinfo {pages} {193--198} (\bibinfo {year}
  {2023})}\BibitemShut {NoStop}%
\bibitem [{\citenamefont {Opatrný}\ \emph {et~al.}(2023)\citenamefont
  {Opatrný}, \citenamefont {Šimon Bräuer}, \citenamefont {Kofman},
  \citenamefont {Misra}, \citenamefont {Meher}, \citenamefont {Firstenberg},
  \citenamefont {Poem},\ and\ \citenamefont {Kurizki}}]{Opatrny2023ScAdv}%
  \BibitemOpen
  \bibfield  {author} {\bibinfo {author} {\bibfnamefont {T.}~\bibnamefont
  {Opatrný}}, \bibinfo {author} {\bibnamefont {Šimon Bräuer}}, \bibinfo
  {author} {\bibfnamefont {A.~G.}\ \bibnamefont {Kofman}}, \bibinfo {author}
  {\bibfnamefont {A.}~\bibnamefont {Misra}}, \bibinfo {author} {\bibfnamefont
  {N.}~\bibnamefont {Meher}}, \bibinfo {author} {\bibfnamefont
  {O.}~\bibnamefont {Firstenberg}}, \bibinfo {author} {\bibfnamefont
  {E.}~\bibnamefont {Poem}},\ and\ \bibinfo {author} {\bibfnamefont
  {G.}~\bibnamefont {Kurizki}},\ }\bibfield  {title} {\enquote {\bibinfo
  {title} {Nonlinear coherent heat machines},}\ }\href
  {https://doi.org/10.1126/sciadv.adf1070} {\bibfield  {journal} {\bibinfo
  {journal} {Science Advances}\ }\textbf {\bibinfo {volume} {9}},\ \bibinfo
  {pages} {eadf1070} (\bibinfo {year} {2023})},\ \Eprint
  {https://arxiv.org/abs/https://www.science.org/doi/pdf/10.1126/sciadv.adf1070}
  {https://www.science.org/doi/pdf/10.1126/sciadv.adf1070} \BibitemShut
  {NoStop}%
\bibitem [{\citenamefont {Meher}, \citenamefont {Opatrny},\ and\ \citenamefont
  {Kurizki}(2024)}]{meher2024thermodynamic}%
  \BibitemOpen
  \bibfield  {author} {\bibinfo {author} {\bibfnamefont {N.}~\bibnamefont
  {Meher}}, \bibinfo {author} {\bibfnamefont {T.}~\bibnamefont {Opatrny}},\
  and\ \bibinfo {author} {\bibfnamefont {G.}~\bibnamefont {Kurizki}},\
  }\bibfield  {title} {\enquote {\bibinfo {title} {Thermodynamic sensing of
  quantum nonlinear noise correlations},}\ }\href@noop {} {\bibfield  {journal}
  {\bibinfo  {journal} {Quantum Science and Technology}\ } (\bibinfo {year}
  {2024})}\BibitemShut {NoStop}%
\bibitem [{\citenamefont {Meher}\ \emph {et~al.}(2024)\citenamefont {Meher},
  \citenamefont {Poem}, \citenamefont {Opatrn{\`y}}, \citenamefont
  {Firstenberg},\ and\ \citenamefont {Kurizki}}]{meher2024supersensitive}%
  \BibitemOpen
  \bibfield  {author} {\bibinfo {author} {\bibfnamefont {N.}~\bibnamefont
  {Meher}}, \bibinfo {author} {\bibfnamefont {E.}~\bibnamefont {Poem}},
  \bibinfo {author} {\bibfnamefont {T.}~\bibnamefont {Opatrn{\`y}}}, \bibinfo
  {author} {\bibfnamefont {O.}~\bibnamefont {Firstenberg}},\ and\ \bibinfo
  {author} {\bibfnamefont {G.}~\bibnamefont {Kurizki}},\ }\bibfield  {title}
  {\enquote {\bibinfo {title} {Supersensitive phase estimation by thermal light
  in a kerr-nonlinear interferometric setup},}\ }\href@noop {} {\bibfield
  {journal} {\bibinfo  {journal} {Physical Review A}\ }\textbf {\bibinfo
  {volume} {110}},\ \bibinfo {pages} {013715} (\bibinfo {year}
  {2024})}\BibitemShut {NoStop}%
\bibitem [{\citenamefont {Opatrn\'y}, \citenamefont {Misra},\ and\
  \citenamefont {Kurizki}(2021)}]{OpatrnyPRL21}%
  \BibitemOpen
  \bibfield  {author} {\bibinfo {author} {\bibfnamefont {T.}~\bibnamefont
  {Opatrn\'y}}, \bibinfo {author} {\bibfnamefont {A.}~\bibnamefont {Misra}},\
  and\ \bibinfo {author} {\bibfnamefont {G.}~\bibnamefont {Kurizki}},\
  }\bibfield  {title} {\enquote {\bibinfo {title} {Work generation from thermal
  noise by quantum phase-sensitive observation},}\ }\href
  {https://doi.org/10.1103/PhysRevLett.127.040602} {\bibfield  {journal}
  {\bibinfo  {journal} {Phys. Rev. Lett.}\ }\textbf {\bibinfo {volume} {127}},\
  \bibinfo {pages} {040602} (\bibinfo {year} {2021})}\BibitemShut {NoStop}%
\bibitem [{\citenamefont {Dasari}\ \emph {et~al.}(2022)\citenamefont {Dasari},
  \citenamefont {Yang}, \citenamefont {Chakrabarti}, \citenamefont {Finkler},
  \citenamefont {Kurizki},\ and\ \citenamefont {Wrachtrup}}]{dasari2022anti}%
  \BibitemOpen
  \bibfield  {author} {\bibinfo {author} {\bibfnamefont {D.~B.~R.}\
  \bibnamefont {Dasari}}, \bibinfo {author} {\bibfnamefont {S.}~\bibnamefont
  {Yang}}, \bibinfo {author} {\bibfnamefont {A.}~\bibnamefont {Chakrabarti}},
  \bibinfo {author} {\bibfnamefont {A.}~\bibnamefont {Finkler}}, \bibinfo
  {author} {\bibfnamefont {G.}~\bibnamefont {Kurizki}},\ and\ \bibinfo {author}
  {\bibfnamefont {J.}~\bibnamefont {Wrachtrup}},\ }\bibfield  {title} {\enquote
  {\bibinfo {title} {Anti-zeno purification of spin baths by quantum probe
  measurements},}\ }\href@noop {} {\bibfield  {journal} {\bibinfo  {journal}
  {Nature communications}\ }\textbf {\bibinfo {volume} {13}},\ \bibinfo {pages}
  {7527} (\bibinfo {year} {2022})}\BibitemShut {NoStop}%
\bibitem [{\citenamefont {Uzdin}\ and\ \citenamefont
  {Rahav}(2018)}]{uzdin2018global}%
  \BibitemOpen
  \bibfield  {author} {\bibinfo {author} {\bibfnamefont {R.}~\bibnamefont
  {Uzdin}}\ and\ \bibinfo {author} {\bibfnamefont {S.}~\bibnamefont {Rahav}},\
  }\bibfield  {title} {\enquote {\bibinfo {title} {Global passivity in
  microscopic thermodynamics},}\ }\href@noop {} {\bibfield  {journal} {\bibinfo
   {journal} {Physical Review X}\ }\textbf {\bibinfo {volume} {8}},\ \bibinfo
  {pages} {021064} (\bibinfo {year} {2018})}\BibitemShut {NoStop}%
\bibitem [{\citenamefont {Kitagawa}\ and\ \citenamefont
  {Ueda}(1993)}]{Kitagawa1993PRA}%
  \BibitemOpen
  \bibfield  {author} {\bibinfo {author} {\bibfnamefont {M.}~\bibnamefont
  {Kitagawa}}\ and\ \bibinfo {author} {\bibfnamefont {M.}~\bibnamefont
  {Ueda}},\ }\bibfield  {title} {\enquote {\bibinfo {title} {Squeezed spin
  states},}\ }\href@noop {} {\bibfield  {journal} {\bibinfo  {journal}
  {Physical Review A}\ }\textbf {\bibinfo {volume} {47}},\ \bibinfo {pages}
  {5138} (\bibinfo {year} {1993})}\BibitemShut {NoStop}%
\bibitem [{\citenamefont {Opatrn{\`y}}(2015{\natexlab{a}})}]{Opatrny2015PRA}%
  \BibitemOpen
  \bibfield  {author} {\bibinfo {author} {\bibfnamefont {T.}~\bibnamefont
  {Opatrn{\`y}}},\ }\bibfield  {title} {\enquote {\bibinfo {title} {Twisting
  tensor and spin squeezing},}\ }\href@noop {} {\bibfield  {journal} {\bibinfo
  {journal} {Physical Review A}\ }\textbf {\bibinfo {volume} {91}},\ \bibinfo
  {pages} {053826} (\bibinfo {year} {2015}{\natexlab{a}})}\BibitemShut
  {NoStop}%
\bibitem [{\citenamefont {Opatrn{\`y}}(2015{\natexlab{b}})}]{Opatrny2015PRA2}%
  \BibitemOpen
  \bibfield  {author} {\bibinfo {author} {\bibfnamefont {T.}~\bibnamefont
  {Opatrn{\`y}}},\ }\bibfield  {title} {\enquote {\bibinfo {title} {Squeezing
  with classical hamiltonians},}\ }\href@noop {} {\bibfield  {journal}
  {\bibinfo  {journal} {Physical Review A}\ }\textbf {\bibinfo {volume} {92}},\
  \bibinfo {pages} {033801} (\bibinfo {year} {2015}{\natexlab{b}})}\BibitemShut
  {NoStop}%
\bibitem [{\citenamefont {Gerry}\ and\ \citenamefont {Knight}(2004)}]{Gerry}%
  \BibitemOpen
  \bibfield  {author} {\bibinfo {author} {\bibfnamefont {C.}~\bibnamefont
  {Gerry}}\ and\ \bibinfo {author} {\bibfnamefont {P.~L.}\ \bibnamefont
  {Knight}},\ }\href@noop {} {\emph {\bibinfo {title} {Introductory Quantum
  Optics}}}\ (\bibinfo  {publisher} {Cambridge University Press},\ \bibinfo
  {year} {2004})\BibitemShut {NoStop}%
\bibitem [{\citenamefont {Carmichael}(1999)}]{Carmichael_BOOK}%
  \BibitemOpen
  \bibfield  {author} {\bibinfo {author} {\bibfnamefont {H.}~\bibnamefont
  {Carmichael}},\ }\href@noop {} {\emph {\bibinfo {title} {Statistical Methods
  in Quantum Optics}}}\ (\bibinfo  {publisher} {Springer},\ \bibinfo {address}
  {Berlin},\ \bibinfo {year} {1999})\BibitemShut {NoStop}%
\bibitem [{\citenamefont {Gardiner}\ and\ \citenamefont
  {Zoller}(2000)}]{Gardiner}%
  \BibitemOpen
  \bibfield  {author} {\bibinfo {author} {\bibfnamefont {C.~W.}\ \bibnamefont
  {Gardiner}}\ and\ \bibinfo {author} {\bibfnamefont {P.}~\bibnamefont
  {Zoller}},\ }\href@noop {} {\emph {\bibinfo {title} {Quantum Noise}}}\
  (\bibinfo  {publisher} {Springer},\ \bibinfo {address} {Berlin},\ \bibinfo
  {year} {2000})\BibitemShut {NoStop}%
\bibitem [{\citenamefont {Scully}\ and\ \citenamefont
  {Zubairy}(2012)}]{ScullyZubairy}%
  \BibitemOpen
  \bibfield  {author} {\bibinfo {author} {\bibfnamefont {M.~O.}\ \bibnamefont
  {Scully}}\ and\ \bibinfo {author} {\bibfnamefont {M.~S.}\ \bibnamefont
  {Zubairy}},\ }\href@noop {} {\emph {\bibinfo {title} {Quantum Optics}}}\
  (\bibinfo  {publisher} {Cambridge University Press},\ \bibinfo {year}
  {2012})\BibitemShut {NoStop}%
\bibitem [{\citenamefont {Afek}, \citenamefont {Ambar},\ and\ \citenamefont
  {Silberberg}(2010)}]{PhysRevLett.104.123602}%
  \BibitemOpen
  \bibfield  {author} {\bibinfo {author} {\bibfnamefont {I.}~\bibnamefont
  {Afek}}, \bibinfo {author} {\bibfnamefont {O.}~\bibnamefont {Ambar}},\ and\
  \bibinfo {author} {\bibfnamefont {Y.}~\bibnamefont {Silberberg}},\ }\bibfield
   {title} {\enquote {\bibinfo {title} {Classical bound for mach-zehnder
  superresolution},}\ }\href {https://doi.org/10.1103/PhysRevLett.104.123602}
  {\bibfield  {journal} {\bibinfo  {journal} {Phys. Rev. Lett.}\ }\textbf
  {\bibinfo {volume} {104}},\ \bibinfo {pages} {123602} (\bibinfo {year}
  {2010})}\BibitemShut {NoStop}%
\bibitem [{\citenamefont {Lang}\ and\ \citenamefont
  {Caves}(2013)}]{lang2013optimal}%
  \BibitemOpen
  \bibfield  {author} {\bibinfo {author} {\bibfnamefont {M.~D.}\ \bibnamefont
  {Lang}}\ and\ \bibinfo {author} {\bibfnamefont {C.~M.}\ \bibnamefont
  {Caves}},\ }\bibfield  {title} {\enquote {\bibinfo {title} {Optimal
  quantum-enhanced interferometry using a laser power source},}\ }\href@noop {}
  {\bibfield  {journal} {\bibinfo  {journal} {Physical review letters}\
  }\textbf {\bibinfo {volume} {111}},\ \bibinfo {pages} {173601} (\bibinfo
  {year} {2013})}\BibitemShut {NoStop}%
\bibitem [{\citenamefont {Luis}(2001)}]{luis2001equivalence}%
  \BibitemOpen
  \bibfield  {author} {\bibinfo {author} {\bibfnamefont {A.}~\bibnamefont
  {Luis}},\ }\bibfield  {title} {\enquote {\bibinfo {title} {Equivalence
  between macroscopic quantum superpositions and maximally entangled states:
  Application to phase-shift detection},}\ }\href@noop {} {\bibfield  {journal}
  {\bibinfo  {journal} {Physical Review A}\ }\textbf {\bibinfo {volume} {64}},\
  \bibinfo {pages} {054102} (\bibinfo {year} {2001})}\BibitemShut {NoStop}%
\bibitem [{\citenamefont {Chekhova}\ and\ \citenamefont
  {Ou}(2016)}]{chekhova2016nonlinear}%
  \BibitemOpen
  \bibfield  {author} {\bibinfo {author} {\bibfnamefont {M.}~\bibnamefont
  {Chekhova}}\ and\ \bibinfo {author} {\bibfnamefont {Z.}~\bibnamefont {Ou}},\
  }\bibfield  {title} {\enquote {\bibinfo {title} {Nonlinear interferometers in
  quantum optics},}\ }\href@noop {} {\bibfield  {journal} {\bibinfo  {journal}
  {Advances in Optics and Photonics}\ }\textbf {\bibinfo {volume} {8}},\
  \bibinfo {pages} {104--155} (\bibinfo {year} {2016})}\BibitemShut {NoStop}%
\bibitem [{\citenamefont {Kitagawa}\ and\ \citenamefont
  {Yamamoto}(1986)}]{kitagawa1986number}%
  \BibitemOpen
  \bibfield  {author} {\bibinfo {author} {\bibfnamefont {M.}~\bibnamefont
  {Kitagawa}}\ and\ \bibinfo {author} {\bibfnamefont {Y.}~\bibnamefont
  {Yamamoto}},\ }\bibfield  {title} {\enquote {\bibinfo {title} {Number-phase
  minimum-uncertainty state with reduced number uncertainty in a kerr nonlinear
  interferometer},}\ }\href@noop {} {\bibfield  {journal} {\bibinfo  {journal}
  {Physical Review A}\ }\textbf {\bibinfo {volume} {34}},\ \bibinfo {pages}
  {3974} (\bibinfo {year} {1986})}\BibitemShut {NoStop}%
\bibitem [{\citenamefont {Cram{\'e}r}(1999)}]{cramer1999mathematical}%
  \BibitemOpen
  \bibfield  {author} {\bibinfo {author} {\bibfnamefont {H.}~\bibnamefont
  {Cram{\'e}r}},\ }\href@noop {} {\emph {\bibinfo {title} {Mathematical methods
  of statistics}}},\ Vol.~\bibinfo {volume} {43}\ (\bibinfo  {publisher}
  {Princeton university press},\ \bibinfo {year} {1999})\BibitemShut {NoStop}%
\bibitem [{\citenamefont {Birrittella}, \citenamefont {Alsing},\ and\
  \citenamefont {Gerry}(2021)}]{birrittella2021parity}%
  \BibitemOpen
  \bibfield  {author} {\bibinfo {author} {\bibfnamefont {R.~J.}\ \bibnamefont
  {Birrittella}}, \bibinfo {author} {\bibfnamefont {P.~M.}\ \bibnamefont
  {Alsing}},\ and\ \bibinfo {author} {\bibfnamefont {C.~C.}\ \bibnamefont
  {Gerry}},\ }\bibfield  {title} {\enquote {\bibinfo {title} {The parity
  operator: Applications in quantum metrology},}\ }\href
  {https://doi.org/10.1116/5.0026148} {\bibfield  {journal} {\bibinfo
  {journal} {AVS Quantum Science}\ }\textbf {\bibinfo {volume} {3}} (\bibinfo
  {year} {2021}),\ 10.1116/5.0026148}\BibitemShut {NoStop}%
\bibitem [{\citenamefont {T{\'o}th}\ and\ \citenamefont
  {Apellaniz}(2014)}]{Toth2014JPA}%
  \BibitemOpen
  \bibfield  {author} {\bibinfo {author} {\bibfnamefont {G.}~\bibnamefont
  {T{\'o}th}}\ and\ \bibinfo {author} {\bibfnamefont {I.}~\bibnamefont
  {Apellaniz}},\ }\bibfield  {title} {\enquote {\bibinfo {title} {Quantum
  metrology from a quantum information science perspective},}\ }\href@noop {}
  {\bibfield  {journal} {\bibinfo  {journal} {Journal of Physics A:
  Mathematical and Theoretical}\ }\textbf {\bibinfo {volume} {47}},\ \bibinfo
  {pages} {424006} (\bibinfo {year} {2014})}\BibitemShut {NoStop}%
\bibitem [{\citenamefont {Ou}(1997)}]{Ou1997PRA}%
  \BibitemOpen
  \bibfield  {author} {\bibinfo {author} {\bibfnamefont {Z.~Y.}\ \bibnamefont
  {Ou}},\ }\bibfield  {title} {\enquote {\bibinfo {title} {Fundamental quantum
  limit in precision phase measurement},}\ }\href
  {https://doi.org/10.1103/PhysRevA.55.2598} {\bibfield  {journal} {\bibinfo
  {journal} {Phys. Rev. A}\ }\textbf {\bibinfo {volume} {55}},\ \bibinfo
  {pages} {2598--2609} (\bibinfo {year} {1997})}\BibitemShut {NoStop}%
\bibitem [{\citenamefont {Monras}(2006)}]{Monras2006PRA}%
  \BibitemOpen
  \bibfield  {author} {\bibinfo {author} {\bibfnamefont {A.}~\bibnamefont
  {Monras}},\ }\bibfield  {title} {\enquote {\bibinfo {title} {Optimal phase
  measurements with pure gaussian states},}\ }\href
  {https://doi.org/10.1103/PhysRevA.73.033821} {\bibfield  {journal} {\bibinfo
  {journal} {Phys. Rev. A}\ }\textbf {\bibinfo {volume} {73}},\ \bibinfo
  {pages} {033821} (\bibinfo {year} {2006})}\BibitemShut {NoStop}%
\bibitem [{\citenamefont {Hofmann}(2009)}]{Hofman2009PRA}%
  \BibitemOpen
  \bibfield  {author} {\bibinfo {author} {\bibfnamefont {H.~F.}\ \bibnamefont
  {Hofmann}},\ }\bibfield  {title} {\enquote {\bibinfo {title} {All
  path-symmetric pure states achieve their maximal phase sensitivity in
  conventional two-path interferometry},}\ }\href
  {https://doi.org/10.1103/PhysRevA.79.033822} {\bibfield  {journal} {\bibinfo
  {journal} {Phys. Rev. A}\ }\textbf {\bibinfo {volume} {79}},\ \bibinfo
  {pages} {033822} (\bibinfo {year} {2009})}\BibitemShut {NoStop}%
\bibitem [{\citenamefont {Pinel}\ \emph {et~al.}(2013)\citenamefont {Pinel},
  \citenamefont {Jian}, \citenamefont {Treps}, \citenamefont {Fabre},\ and\
  \citenamefont {Braun}}]{Pinel2013PRA}%
  \BibitemOpen
  \bibfield  {author} {\bibinfo {author} {\bibfnamefont {O.}~\bibnamefont
  {Pinel}}, \bibinfo {author} {\bibfnamefont {P.}~\bibnamefont {Jian}},
  \bibinfo {author} {\bibfnamefont {N.}~\bibnamefont {Treps}}, \bibinfo
  {author} {\bibfnamefont {C.}~\bibnamefont {Fabre}},\ and\ \bibinfo {author}
  {\bibfnamefont {D.}~\bibnamefont {Braun}},\ }\bibfield  {title} {\enquote
  {\bibinfo {title} {Quantum parameter estimation using general single-mode
  gaussian states},}\ }\href {https://doi.org/10.1103/PhysRevA.88.040102}
  {\bibfield  {journal} {\bibinfo  {journal} {Phys. Rev. A}\ }\textbf {\bibinfo
  {volume} {88}},\ \bibinfo {pages} {040102} (\bibinfo {year}
  {2013})}\BibitemShut {NoStop}%
\bibitem [{\citenamefont {Boixo}\ \emph {et~al.}(2007)\citenamefont {Boixo},
  \citenamefont {Flammia}, \citenamefont {Caves},\ and\ \citenamefont
  {Geremia}}]{Boixo2007PRL}%
  \BibitemOpen
  \bibfield  {author} {\bibinfo {author} {\bibfnamefont {S.}~\bibnamefont
  {Boixo}}, \bibinfo {author} {\bibfnamefont {S.~T.}\ \bibnamefont {Flammia}},
  \bibinfo {author} {\bibfnamefont {C.~M.}\ \bibnamefont {Caves}},\ and\
  \bibinfo {author} {\bibfnamefont {J.}~\bibnamefont {Geremia}},\ }\bibfield
  {title} {\enquote {\bibinfo {title} {Generalized limits for single-parameter
  quantum estimation},}\ }\href {https://doi.org/10.1103/PhysRevLett.98.090401}
  {\bibfield  {journal} {\bibinfo  {journal} {Phys. Rev. Lett.}\ }\textbf
  {\bibinfo {volume} {98}},\ \bibinfo {pages} {090401} (\bibinfo {year}
  {2007})}\BibitemShut {NoStop}%
\bibitem [{\citenamefont {Giovannetti}, \citenamefont {Lloyd},\ and\
  \citenamefont {Maccone}(2012)}]{Giovannetti2012PRL}%
  \BibitemOpen
  \bibfield  {author} {\bibinfo {author} {\bibfnamefont {V.}~\bibnamefont
  {Giovannetti}}, \bibinfo {author} {\bibfnamefont {S.}~\bibnamefont {Lloyd}},\
  and\ \bibinfo {author} {\bibfnamefont {L.}~\bibnamefont {Maccone}},\
  }\bibfield  {title} {\enquote {\bibinfo {title} {Quantum measurement bounds
  beyond the uncertainty relations},}\ }\href
  {https://doi.org/10.1103/PhysRevLett.108.260405} {\bibfield  {journal}
  {\bibinfo  {journal} {Phys. Rev. Lett.}\ }\textbf {\bibinfo {volume} {108}},\
  \bibinfo {pages} {260405} (\bibinfo {year} {2012})}\BibitemShut {NoStop}%
\bibitem [{\citenamefont {Chuang}\ and\ \citenamefont
  {Nielsen}(1997)}]{Chuang1997JModOpt}%
  \BibitemOpen
  \bibfield  {author} {\bibinfo {author} {\bibfnamefont {I.~L.}\ \bibnamefont
  {Chuang}}\ and\ \bibinfo {author} {\bibfnamefont {M.~A.}\ \bibnamefont
  {Nielsen}},\ }\bibfield  {title} {\enquote {\bibinfo {title} {Prescription
  for experimental determination of the dynamics of a quantum black box},}\
  }\href@noop {} {\bibfield  {journal} {\bibinfo  {journal} {Journal of Modern
  Optics}\ }\textbf {\bibinfo {volume} {44}},\ \bibinfo {pages} {2455--2467}
  (\bibinfo {year} {1997})}\BibitemShut {NoStop}%
\bibitem [{\citenamefont {Goldman}\ and\ \citenamefont
  {Dalibard}(2014)}]{Goldman2014PRX}%
  \BibitemOpen
  \bibfield  {author} {\bibinfo {author} {\bibfnamefont {N.}~\bibnamefont
  {Goldman}}\ and\ \bibinfo {author} {\bibfnamefont {J.}~\bibnamefont
  {Dalibard}},\ }\bibfield  {title} {\enquote {\bibinfo {title} {Periodically
  driven quantum systems: Effective hamiltonians and engineered gauge
  fields},}\ }\href {https://doi.org/10.1103/PhysRevX.4.031027} {\bibfield
  {journal} {\bibinfo  {journal} {Phys. Rev. X}\ }\textbf {\bibinfo {volume}
  {4}},\ \bibinfo {pages} {031027} (\bibinfo {year} {2014})}\BibitemShut
  {NoStop}%
\bibitem [{\citenamefont {Del~Pino}\ and\ \citenamefont
  {Zilberberg}(2023)}]{del2023dynamical}%
  \BibitemOpen
  \bibfield  {author} {\bibinfo {author} {\bibfnamefont {J.}~\bibnamefont
  {Del~Pino}}\ and\ \bibinfo {author} {\bibfnamefont {O.}~\bibnamefont
  {Zilberberg}},\ }\bibfield  {title} {\enquote {\bibinfo {title} {Dynamical
  gauge fields with bosonic codes},}\ }\href@noop {} {\bibfield  {journal}
  {\bibinfo  {journal} {Physical Review Letters}\ }\textbf {\bibinfo {volume}
  {130}},\ \bibinfo {pages} {171901} (\bibinfo {year} {2023})}\BibitemShut
  {NoStop}%
\bibitem [{\citenamefont {Mohseni}, \citenamefont {Rezakhani},\ and\
  \citenamefont {Lidar}(2008)}]{Mohseni2008PRA}%
  \BibitemOpen
  \bibfield  {author} {\bibinfo {author} {\bibfnamefont {M.}~\bibnamefont
  {Mohseni}}, \bibinfo {author} {\bibfnamefont {A.~T.}\ \bibnamefont
  {Rezakhani}},\ and\ \bibinfo {author} {\bibfnamefont {D.~A.}\ \bibnamefont
  {Lidar}},\ }\bibfield  {title} {\enquote {\bibinfo {title} {Quantum-process
  tomography: Resource analysis of different strategies},}\ }\href
  {https://doi.org/10.1103/PhysRevA.77.032322} {\bibfield  {journal} {\bibinfo
  {journal} {Phys. Rev. A}\ }\textbf {\bibinfo {volume} {77}},\ \bibinfo
  {pages} {032322} (\bibinfo {year} {2008})}\BibitemShut {NoStop}%
\bibitem [{\citenamefont {Lvovsky}\ and\ \citenamefont
  {Raymer}(2009)}]{Lvovsky2009RMP}%
  \BibitemOpen
  \bibfield  {author} {\bibinfo {author} {\bibfnamefont {A.~I.}\ \bibnamefont
  {Lvovsky}}\ and\ \bibinfo {author} {\bibfnamefont {M.~G.}\ \bibnamefont
  {Raymer}},\ }\bibfield  {title} {\enquote {\bibinfo {title}
  {Continuous-variable optical quantum-state tomography},}\ }\href
  {https://doi.org/10.1103/RevModPhys.81.299} {\bibfield  {journal} {\bibinfo
  {journal} {Rev. Mod. Phys.}\ }\textbf {\bibinfo {volume} {81}},\ \bibinfo
  {pages} {299--332} (\bibinfo {year} {2009})}\BibitemShut {NoStop}%
\bibitem [{\citenamefont {Harel}\ and\ \citenamefont
  {Kurizki}(1996)}]{harel1996fock}%
  \BibitemOpen
  \bibfield  {author} {\bibinfo {author} {\bibfnamefont {G.}~\bibnamefont
  {Harel}}\ and\ \bibinfo {author} {\bibfnamefont {G.}~\bibnamefont
  {Kurizki}},\ }\bibfield  {title} {\enquote {\bibinfo {title} {Fock-state
  preparation from thermal cavity fields by measurements on resonant atoms},}\
  }\href@noop {} {\bibfield  {journal} {\bibinfo  {journal} {Physical Review
  A}\ }\textbf {\bibinfo {volume} {54}},\ \bibinfo {pages} {5410} (\bibinfo
  {year} {1996})}\BibitemShut {NoStop}%
\bibitem [{\citenamefont {Brune}\ \emph {et~al.}(1996)\citenamefont {Brune},
  \citenamefont {Hagley}, \citenamefont {Dreyer}, \citenamefont {Maitre},
  \citenamefont {Maali}, \citenamefont {Wunderlich}, \citenamefont {Raimond},\
  and\ \citenamefont {Haroche}}]{brune1996observing}%
  \BibitemOpen
  \bibfield  {author} {\bibinfo {author} {\bibfnamefont {M.}~\bibnamefont
  {Brune}}, \bibinfo {author} {\bibfnamefont {E.}~\bibnamefont {Hagley}},
  \bibinfo {author} {\bibfnamefont {J.}~\bibnamefont {Dreyer}}, \bibinfo
  {author} {\bibfnamefont {X.}~\bibnamefont {Maitre}}, \bibinfo {author}
  {\bibfnamefont {A.}~\bibnamefont {Maali}}, \bibinfo {author} {\bibfnamefont
  {C.}~\bibnamefont {Wunderlich}}, \bibinfo {author} {\bibfnamefont {J.-M.}\
  \bibnamefont {Raimond}},\ and\ \bibinfo {author} {\bibfnamefont
  {S.}~\bibnamefont {Haroche}},\ }\bibfield  {title} {\enquote {\bibinfo
  {title} {Observing the progressive decoherence of the “meter” in a
  quantum measurement},}\ }\href@noop {} {\bibfield  {journal} {\bibinfo
  {journal} {Physical review letters}\ }\textbf {\bibinfo {volume} {77}},\
  \bibinfo {pages} {4887} (\bibinfo {year} {1996})}\BibitemShut {NoStop}%
\bibitem [{\citenamefont {Jaynes}\ and\ \citenamefont
  {Cummings}(1963)}]{jaynes1963comparison}%
  \BibitemOpen
  \bibfield  {author} {\bibinfo {author} {\bibfnamefont {E.~T.}\ \bibnamefont
  {Jaynes}}\ and\ \bibinfo {author} {\bibfnamefont {F.~W.}\ \bibnamefont
  {Cummings}},\ }\bibfield  {title} {\enquote {\bibinfo {title} {Comparison of
  quantum and semiclassical radiation theories with application to the beam
  maser},}\ }\href@noop {} {\bibfield  {journal} {\bibinfo  {journal}
  {Proceedings of the IEEE}\ }\textbf {\bibinfo {volume} {51}},\ \bibinfo
  {pages} {89--109} (\bibinfo {year} {1963})}\BibitemShut {NoStop}%
\bibitem [{\citenamefont {Harel}\ \emph {et~al.}(1996)\citenamefont {Harel},
  \citenamefont {Kurizki}, \citenamefont {McIver},\ and\ \citenamefont
  {Coutsias}}]{harel1996optimized}%
  \BibitemOpen
  \bibfield  {author} {\bibinfo {author} {\bibfnamefont {G.}~\bibnamefont
  {Harel}}, \bibinfo {author} {\bibfnamefont {G.}~\bibnamefont {Kurizki}},
  \bibinfo {author} {\bibfnamefont {J.}~\bibnamefont {McIver}},\ and\ \bibinfo
  {author} {\bibfnamefont {E.}~\bibnamefont {Coutsias}},\ }\bibfield  {title}
  {\enquote {\bibinfo {title} {Optimized preparation of quantum states by
  conditional measurements},}\ }\href@noop {} {\bibfield  {journal} {\bibinfo
  {journal} {Physical Review A}\ }\textbf {\bibinfo {volume} {53}},\ \bibinfo
  {pages} {4534} (\bibinfo {year} {1996})}\BibitemShut {NoStop}%
\bibitem [{\citenamefont {Garraway}\ \emph {et~al.}(1994)\citenamefont
  {Garraway}, \citenamefont {Sherman}, \citenamefont {Moya-Cessa},
  \citenamefont {Knight},\ and\ \citenamefont {Kurizki}}]{Garraway1994PRA}%
  \BibitemOpen
  \bibfield  {author} {\bibinfo {author} {\bibfnamefont {B.~M.}\ \bibnamefont
  {Garraway}}, \bibinfo {author} {\bibfnamefont {B.}~\bibnamefont {Sherman}},
  \bibinfo {author} {\bibfnamefont {H.}~\bibnamefont {Moya-Cessa}}, \bibinfo
  {author} {\bibfnamefont {P.~L.}\ \bibnamefont {Knight}},\ and\ \bibinfo
  {author} {\bibfnamefont {G.}~\bibnamefont {Kurizki}},\ }\bibfield  {title}
  {\enquote {\bibinfo {title} {Generation and detection of nonclassical field
  states by conditional measurements following two-photon resonant
  interactions},}\ }\href {https://doi.org/10.1103/PhysRevA.49.535} {\bibfield
  {journal} {\bibinfo  {journal} {Phys. Rev. A}\ }\textbf {\bibinfo {volume}
  {49}},\ \bibinfo {pages} {535--547} (\bibinfo {year} {1994})}\BibitemShut
  {NoStop}%
\bibitem [{\citenamefont {Kofman}\ and\ \citenamefont
  {Kurizki}(2000)}]{kofman2000acceleration}%
  \BibitemOpen
  \bibfield  {author} {\bibinfo {author} {\bibfnamefont {A.}~\bibnamefont
  {Kofman}}\ and\ \bibinfo {author} {\bibfnamefont {G.}~\bibnamefont
  {Kurizki}},\ }\bibfield  {title} {\enquote {\bibinfo {title} {Acceleration of
  quantum decay processes by frequent observations},}\ }\href@noop {}
  {\bibfield  {journal} {\bibinfo  {journal} {Nature}\ }\textbf {\bibinfo
  {volume} {405}},\ \bibinfo {pages} {546--550} (\bibinfo {year}
  {2000})}\BibitemShut {NoStop}%
\bibitem [{\citenamefont {Kofman}\ and\ \citenamefont
  {Kurizki}(2001)}]{kofman2001universal}%
  \BibitemOpen
  \bibfield  {author} {\bibinfo {author} {\bibfnamefont {A.}~\bibnamefont
  {Kofman}}\ and\ \bibinfo {author} {\bibfnamefont {G.}~\bibnamefont
  {Kurizki}},\ }\bibfield  {title} {\enquote {\bibinfo {title} {Universal
  dynamical control of quantum mechanical decay: modulation of the coupling to
  the continuum},}\ }\href@noop {} {\bibfield  {journal} {\bibinfo  {journal}
  {Physical review letters}\ }\textbf {\bibinfo {volume} {87}},\ \bibinfo
  {pages} {270405} (\bibinfo {year} {2001})}\BibitemShut {NoStop}%
\bibitem [{\citenamefont {Virz\`{\i}}\ \emph {et~al.}(2022)\citenamefont
  {Virz\`{\i}}, \citenamefont {Avella}, \citenamefont {Piacentini},
  \citenamefont {Gramegna}, \citenamefont {Opatrn\'y}, \citenamefont {Kofman},
  \citenamefont {Kurizki}, \citenamefont {Gherardini}, \citenamefont {Caruso},
  \citenamefont {Degiovanni},\ and\ \citenamefont {Genovese}}]{VirziPRL2022}%
  \BibitemOpen
  \bibfield  {author} {\bibinfo {author} {\bibfnamefont {S.}~\bibnamefont
  {Virz\`{\i}}}, \bibinfo {author} {\bibfnamefont {A.}~\bibnamefont {Avella}},
  \bibinfo {author} {\bibfnamefont {F.}~\bibnamefont {Piacentini}}, \bibinfo
  {author} {\bibfnamefont {M.}~\bibnamefont {Gramegna}}, \bibinfo {author}
  {\bibfnamefont {T.~c.~v.}\ \bibnamefont {Opatrn\'y}}, \bibinfo {author}
  {\bibfnamefont {A.~G.}\ \bibnamefont {Kofman}}, \bibinfo {author}
  {\bibfnamefont {G.}~\bibnamefont {Kurizki}}, \bibinfo {author} {\bibfnamefont
  {S.}~\bibnamefont {Gherardini}}, \bibinfo {author} {\bibfnamefont
  {F.}~\bibnamefont {Caruso}}, \bibinfo {author} {\bibfnamefont {I.~P.}\
  \bibnamefont {Degiovanni}},\ and\ \bibinfo {author} {\bibfnamefont
  {M.}~\bibnamefont {Genovese}},\ }\bibfield  {title} {\enquote {\bibinfo
  {title} {Quantum zeno and anti-zeno probes of noise correlations in photon
  polarization},}\ }\href {https://doi.org/10.1103/PhysRevLett.129.030401}
  {\bibfield  {journal} {\bibinfo  {journal} {Phys. Rev. Lett.}\ }\textbf
  {\bibinfo {volume} {129}},\ \bibinfo {pages} {030401} (\bibinfo {year}
  {2022})}\BibitemShut {NoStop}%
\bibitem [{\citenamefont {Kofman}, \citenamefont {Kurizki},\ and\ \citenamefont
  {Opatrn{\`y}}(2001)}]{kofman2001zeno}%
  \BibitemOpen
  \bibfield  {author} {\bibinfo {author} {\bibfnamefont {A.}~\bibnamefont
  {Kofman}}, \bibinfo {author} {\bibfnamefont {G.}~\bibnamefont {Kurizki}},\
  and\ \bibinfo {author} {\bibfnamefont {T.}~\bibnamefont {Opatrn{\`y}}},\
  }\bibfield  {title} {\enquote {\bibinfo {title} {Zeno and anti-zeno effects
  for photon polarization dephasing},}\ }\href@noop {} {\bibfield  {journal}
  {\bibinfo  {journal} {Physical Review A}\ }\textbf {\bibinfo {volume} {63}},\
  \bibinfo {pages} {042108} (\bibinfo {year} {2001})}\BibitemShut {NoStop}%
\bibitem [{\citenamefont {Misra}\ \emph {et~al.}(2024)\citenamefont {Misra},
  \citenamefont {Chattopadhyay}, \citenamefont {Svidzinsky}, \citenamefont
  {Scully},\ and\ \citenamefont {Kurizki}}]{misra2024black}%
  \BibitemOpen
  \bibfield  {author} {\bibinfo {author} {\bibfnamefont {A.}~\bibnamefont
  {Misra}}, \bibinfo {author} {\bibfnamefont {P.}~\bibnamefont
  {Chattopadhyay}}, \bibinfo {author} {\bibfnamefont {A.}~\bibnamefont
  {Svidzinsky}}, \bibinfo {author} {\bibfnamefont {M.~O.}\ \bibnamefont
  {Scully}},\ and\ \bibinfo {author} {\bibfnamefont {G.}~\bibnamefont
  {Kurizki}},\ }\bibfield  {title} {\enquote {\bibinfo {title} {Black-hole
  powered quantum coherent amplifier},}\ }\href@noop {} {\bibfield  {journal}
  {\bibinfo  {journal} {npj Quantum Information}\ }\textbf {\bibinfo {volume}
  {10}},\ \bibinfo {pages} {34} (\bibinfo {year} {2024})}\BibitemShut {NoStop}%
\bibitem [{\citenamefont {Friedler}\ \emph {et~al.}(2005)\citenamefont
  {Friedler}, \citenamefont {Petrosyan}, \citenamefont {Fleischhauer},\ and\
  \citenamefont {Kurizki}}]{Friedler}%
  \BibitemOpen
  \bibfield  {author} {\bibinfo {author} {\bibfnamefont {I.}~\bibnamefont
  {Friedler}}, \bibinfo {author} {\bibfnamefont {D.}~\bibnamefont {Petrosyan}},
  \bibinfo {author} {\bibfnamefont {M.}~\bibnamefont {Fleischhauer}},\ and\
  \bibinfo {author} {\bibfnamefont {G.}~\bibnamefont {Kurizki}},\ }\bibfield
  {title} {\enquote {\bibinfo {title} {Long-range interactions and entanglement
  of slow single-photon pulses},}\ }\href
  {https://doi.org/10.1103/PhysRevA.72.043803} {\bibfield  {journal} {\bibinfo
  {journal} {Phys. Rev. A}\ }\textbf {\bibinfo {volume} {72}},\ \bibinfo
  {pages} {043803} (\bibinfo {year} {2005})}\BibitemShut {NoStop}%
\bibitem [{\citenamefont {Imamo\ifmmode~\bar{g}\else \={g}\fi{}lu}\ \emph
  {et~al.}(1997)\citenamefont {Imamo\ifmmode~\bar{g}\else \={g}\fi{}lu},
  \citenamefont {Schmidt}, \citenamefont {Woods},\ and\ \citenamefont
  {Deutsch}}]{Imamoglu1997PRL}%
  \BibitemOpen
  \bibfield  {author} {\bibinfo {author} {\bibfnamefont {A.}~\bibnamefont
  {Imamo\ifmmode~\bar{g}\else \={g}\fi{}lu}}, \bibinfo {author} {\bibfnamefont
  {H.}~\bibnamefont {Schmidt}}, \bibinfo {author} {\bibfnamefont
  {G.}~\bibnamefont {Woods}},\ and\ \bibinfo {author} {\bibfnamefont
  {M.}~\bibnamefont {Deutsch}},\ }\bibfield  {title} {\enquote {\bibinfo
  {title} {Strongly interacting photons in a nonlinear cavity},}\ }\href
  {https://doi.org/10.1103/PhysRevLett.79.1467} {\bibfield  {journal} {\bibinfo
   {journal} {Phys. Rev. Lett.}\ }\textbf {\bibinfo {volume} {79}},\ \bibinfo
  {pages} {1467--1470} (\bibinfo {year} {1997})}\BibitemShut {NoStop}%
\bibitem [{\citenamefont {Karnieli}\ \emph {et~al.}(2024)\citenamefont
  {Karnieli}, \citenamefont {Roques-Carmes}, \citenamefont {Rivera},\ and\
  \citenamefont {Fan}}]{karnieli2024strong}%
  \BibitemOpen
  \bibfield  {author} {\bibinfo {author} {\bibfnamefont {A.}~\bibnamefont
  {Karnieli}}, \bibinfo {author} {\bibfnamefont {C.}~\bibnamefont
  {Roques-Carmes}}, \bibinfo {author} {\bibfnamefont {N.}~\bibnamefont
  {Rivera}},\ and\ \bibinfo {author} {\bibfnamefont {S.}~\bibnamefont {Fan}},\
  }\bibfield  {title} {\enquote {\bibinfo {title} {Strong coupling and
  single-photon nonlinearity in free-electron quantum optics},}\ }\href@noop {}
  {\bibfield  {journal} {\bibinfo  {journal} {ACS Photonics}\ }\textbf
  {\bibinfo {volume} {11}},\ \bibinfo {pages} {3401--3411} (\bibinfo {year}
  {2024})}\BibitemShut {NoStop}%
\bibitem [{\citenamefont {Bhaktavatsala~Rao}, \citenamefont {Bar-Gill},\ and\
  \citenamefont {Kurizki}(2011)}]{DBDRao2011PRL}%
  \BibitemOpen
  \bibfield  {author} {\bibinfo {author} {\bibfnamefont {D.~D.}\ \bibnamefont
  {Bhaktavatsala~Rao}}, \bibinfo {author} {\bibfnamefont {N.}~\bibnamefont
  {Bar-Gill}},\ and\ \bibinfo {author} {\bibfnamefont {G.}~\bibnamefont
  {Kurizki}},\ }\bibfield  {title} {\enquote {\bibinfo {title} {Generation of
  macroscopic superpositions of quantum states by linear coupling to a bath},}\
  }\href {https://doi.org/10.1103/PhysRevLett.106.010404} {\bibfield  {journal}
  {\bibinfo  {journal} {Phys. Rev. Lett.}\ }\textbf {\bibinfo {volume} {106}},\
  \bibinfo {pages} {010404} (\bibinfo {year} {2011})}\BibitemShut {NoStop}%
\end{thebibliography}
%

\end{document}